\documentclass[sigconf]{acmart}

\usepackage{listings}
\usepackage{xcolor}
\usepackage{textcomp}
\usepackage{graphicx}
\usepackage{caption}
\usepackage{subcaption}
\usepackage{todonotes}
\usepackage{multirow}
\usepackage{booktabs}
\usepackage{xspace}
\usepackage{paralist}
\usepackage{amsmath}
\usepackage{enumitem}
\usepackage{tabulary}
\usepackage{prettyref}
\usepackage{verbatim}
\usepackage{balance}

\usepackage{pifont}

\newcounter{dn}

\lstdefinestyle{cstyle}{
  basicstyle=\footnotesize\ttfamily,
  keywordstyle=\color{black!85}\bfseries,
  keywordstyle=[2]\color{black!85}\bfseries\emph,
  showstringspaces=false,
  language={C},
  breaklines=false,
  mathescape=true,
  escapechar={@}
}
\lstdefinestyle{inline}{
  style=cstyle,
  mathescape=false,
  breaklines=true,
  keywordstyle=,            
  keywordstyle=[2],
  extendedchars=true,
  basicstyle=\ttfamily\small
}

\lstdefinelanguage
   [x64]{Assembler}     
   [x86masm]{Assembler} 
   {morekeywords={CDQE,CQO,CMPSQ,CMPXCHG16B,JRCXZ,LODSQ,MOVSXD, %
                  POPFQ,PUSHFQ,SCASQ,STOSQ,IRETQ,RDTSCP,SWAPGS, %
                  LFENCE,CPUID,CMP,
                  rax,rdx,rcx,rbx,rsi,rdi,rsp,rbp, %
                  r8,r8d,r8w,r8b,r9,r9d,r9w,r9b, %
                  r10,r10d,r10w,r10b,r11,r11d,r11w,r11b, %
                  r12,r12d,r12w,r12b,r13,r13d,r13w,r13b, %
                  r14,r14d,r14w,r14b,r15,r15d,r15w,r15b, xmm0, movq}}

\newcommand{\smother}{\textsc{SMoTher}\xspace}
\newcommand{\smotherspectre}{\textsc{SMoTherSpectre}\xspace}

\newrefformat{cha}{\hyperref[#1]{Chapter~\ref*{#1}}}
\newrefformat{sec}{\hyperref[#1]{Section~\ref*{#1}}}
\newrefformat{sub}{\hyperref[#1]{Section~\ref*{#1}}}
\newrefformat{tab}{\hyperref[#1]{Table~\ref*{#1}}}
\newrefformat{fig}{\hyperref[#1]{Figure~\ref*{#1}}}
\newrefformat{line}{\hyperref[#1]{line~\ref*{#1}}}
\newrefformat{lst}{\hyperref[#1]{Listing~\ref*{#1}}}
\newrefformat{pat}{\hyperref[#1]{Patch~\ref*{#1}}}
\newrefformat{alg}{\hyperref[#1]{Algorithm~\ref*{#1}}}

\newcommand{\Code}[1]{\lstinline[style=inline,breaklines=false]@#1@}

\let\paragraph\relax
\newcommand{\paragraph}[1]{\textbf{#1.}}

\newenvironment{minilisting}
  {\minipage[b]{\linewidth}\verbatim}
  {\endverbatim\endminipage}

\newcommand{\aref}[1]{\hyperref[#1]{Appendix~\ref{#1}}}

\copyrightyear{2019}
\acmYear{2019}
\setcopyright{acmlicensed}
\acmConference[CCS '19] {2019 ACM SIGSAC Conference on Computer \& Communications Security}{November 11--15, 2019}{London, United Kingdom}
\acmBooktitle{2019 ACM SIGSAC Conference on Computer \& Communications Security (CCS '19), November 11--15, 2019, London, United Kingdom}
\acmPrice{15.00}
\acmDOI{10.1145/3319535.3363194}
\acmISBN{978-1-4503-6747-9/19/11}

\begin{document}

\fancyhead{}

\hypersetup{
  pdfauthor = {},
  pdftitle = {SMoTherSpectre: exploiting speculative execution through port contention},
  pdfsubject = {},
  pdfkeywords = {},
  backref = true,
  pagebackref  = true,
  colorlinks = true,
  linkcolor = black,
  anchorcolor = black,
  citecolor = black,
  filecolor = black,
  urlcolor = black,
  pagecolor= black
}

\title{SMoTherSpectre: Exploiting Speculative Execution \\through Port Contention}

\author{Atri Bhattacharyya }
\authornote{firstname.surname@epfl.ch}
\affiliation{
  \institution{EPFL}
}
\author {Alexandra Sandulescu }
\authornote{\{asa, eug, aso, kur\}@zurich.ibm.com}
\affiliation{
  \institution{IBM Research -- Zurich}
}
\author {Matthias Neugschwandtner}
\authornotemark[2]
\affiliation{
  \institution{IBM Research -- Zurich}
}
\author {Alessandro Sorniotti }
\authornotemark[2]
\affiliation{
  \institution{IBM Research -- Zurich}
}
\author {Babak Falsafi}
\authornotemark[1]
\affiliation{
  \institution{EPFL}
}
\author {Mathias Payer}
\authornotemark[1]
\affiliation{
  \institution{EPFL}
}
\author {Anil Kurmus }
\authornotemark[2]
\affiliation{
  \institution{IBM Research -- Zurich}
}

\begin{abstract}
Spectre, Meltdown, and related attacks have demonstrated that kernels,
hypervisors, trusted execution environments, and browsers are prone to
information disclosure through micro-architectural weaknesses.  However, it
remains unclear as to what extent other applications, in particular those that do
not load attacker-provided code, may be impacted. It also remains unclear as to
what extent these attacks are reliant on cache-based side channels.

We introduce \smotherspectre, a \emph{speculative code-reuse attack} that leverages
port-contention in simultaneously multi-threaded processors (\smother) as a side
channel to leak information from a victim process. 
\smother is a fine-grained side channel that detects contention based on a
single victim instruction. To discover real-world gadgets, we describe a
methodology and build a tool that locates \smother-gadgets in popular libraries.
In an evaluation on \Code{glibc}, we found hundreds of gadgets that can be
used to leak information.  Finally, we demonstrate proof-of-concept
attacks against the OpenSSH server, creating oracles for determining
four host key bits, and against an application performing encryption 
using the OpenSSL library, creating an oracle which can
differentiate a bit of the plaintext through gadgets in
\Code{libcrypto} and \Code{glibc}.

\end{abstract}



\maketitle

\section{Introduction}
\label{sec:intro}

Spectre~\cite{kocher18oakland, koruyeh18woot, schwarz18arxiv} and 
Meltdown~\cite{lipp18sec} form a new class of micro-architectural attacks. These 
attacks leverage weaknesses in speculative execution (Spectre) or separation 
between privileged and unprivileged code (Meltdown) to leave micro-architectural 
traces~\cite{canella18arxiv}. Both Spectre and Meltdown leverage a side channel 
based on the memory architecture to leak data from the address space of a 
target (e.g. from another process or from the kernel).

While micro-architectural side channels were known before the discovery of
Meltdown and Spectre, their applicability was mostly limited to targets applying
data-dependent control flow patterns or memory accesses.  In this older class of
vulnerabilities, an attacker would observe the micro-architectural changes to
shared resources caused by the execution of a victim. For example, in a
cache-based attack, the adversary would prime the cache, let the victim execute,
and then detect which locations have been evicted from the cache. Such a side
channel leaks addresses and allows the adversary to learn information from
data-dependent execution. An effective mitigation strategy is to eliminate
data-dependent control flow over sensitive data, such as cryptographic material.

In contrast, Spectre and Meltdown render this class of attacks generic and 
significantly harder to mitigate through software changes only. The side channel
is now used \emph{indirectly}, in a way that -- crucially -- does not rely on
poor choices in the development of the target application. In Spectre, for
instance, the attacker first primes the speculation engine (e.g., by preparing
the branch target buffers) as well as the cache-based side channel; the victim
then misspeculates at an attacker-controlled location and thereby leaks
information~\cite{canella18arxiv}.  The attacker can then read out the
cache-based side channel. In light of these new attack vectors,
architectural, system-wide defenses such as Kernel Page-Table 
Isolation~\cite{gruss2017kaslr}, retpolines~\cite{turner2018retpoline}, or 
microcode updates must be rolled out to protect the
system against attacks. One proposed microarchitectural defense is to revert all
side effects of speculative execution~\cite{khasawneh2018safespec}.

\sloppy
One mitigating factor is that so far, with the exception 
of Netspectre-AVX~\cite{schwarz18arxiv}, all existing attacks rely on side channels that are
invariably cache-based to read out information.  This in turn requires the
presence of specific gadgets in the victim, which are often hard to find.
Consider the example of Branch Target Injection (BTI), the technique used in
Spectre v2~\cite{kocher18oakland}: in the initial exploit, no suitable gadget
was identified in the kernel. The attack was successful because it redirected
speculative control flow to externally provided code, in the form of eBPF kernel
code. This observation justifies why mitigations such as retpoline are not
employed at large by user-space programs.

In this paper, we show that speculation attacks (e.g., through branch target 
injection) can leak arbitrary secrets from generic user-space programs through a 
side channel that is not based on the memory architecture. In particular, we 
show that branch target injection can be used on existing program code, without 
requiring the injection of attacker code.  To this end, we first show that port 
contention can be used as a powerful side channel when executing with 
simultaneous multi-threading (\smother).  We then exploit port contention as a 
side channel to transmit information during speculative execution 
(\smotherspectre). This shows that, because the transmission occurs before 
speculative execution ends, reverting side effects of speculative execution 
would not be sufficient as a defense. Finally, we show how 
suitable portions of code can be found in target binaries automatically.

Other related work has looked at execution-unit-sharing as a side
channel~\cite{wang06smt, aciicmez2007cheap, fogh16shotgun, aldaya18portsmash}.
Portsmash~\cite{aldaya18portsmash}, concurrently developed to our work,
demonstrates that port sharing leaks code access patterns and successfully
extracts secrets from a known vulnerable version of OpenSSL. We are, however, the
first to characterize this side channel and leverage it for a speculative
execution attack, providing a full working proof of concept that leaks data from
an up-to-date OpenSSL version. Further, we attack the
OpenSSH server, leaking bits from the host's RSA key.

This paper makes the following contributions:
\begin{itemize}
\item A precise characterization of the port-contention side channel 
(\smother); 
\item A speculative execution attack (\smotherspectre) that demonstrates the 
suitability of non-cache-based side channels to leak information. We show an 
end-to-end attack using speculation based on BTI by combining it with the port 
contention side channel;
\item An automated technique to find target speculative gadgets in programs; and
\item Real world attacks where we target BTI gadgets in the OpenSSH server
and in the latest version of
OpenSSL, along with a \smother gadget from the libc.
\end{itemize}

\section{Background} 

The work in this paper relies on the complex interplay between software and
hardware. In the following, we provide the background information
necessary to understand \smother and \smotherspectre.

\begin{figure}[]
    \centering
      \includegraphics[width=0.8\linewidth]{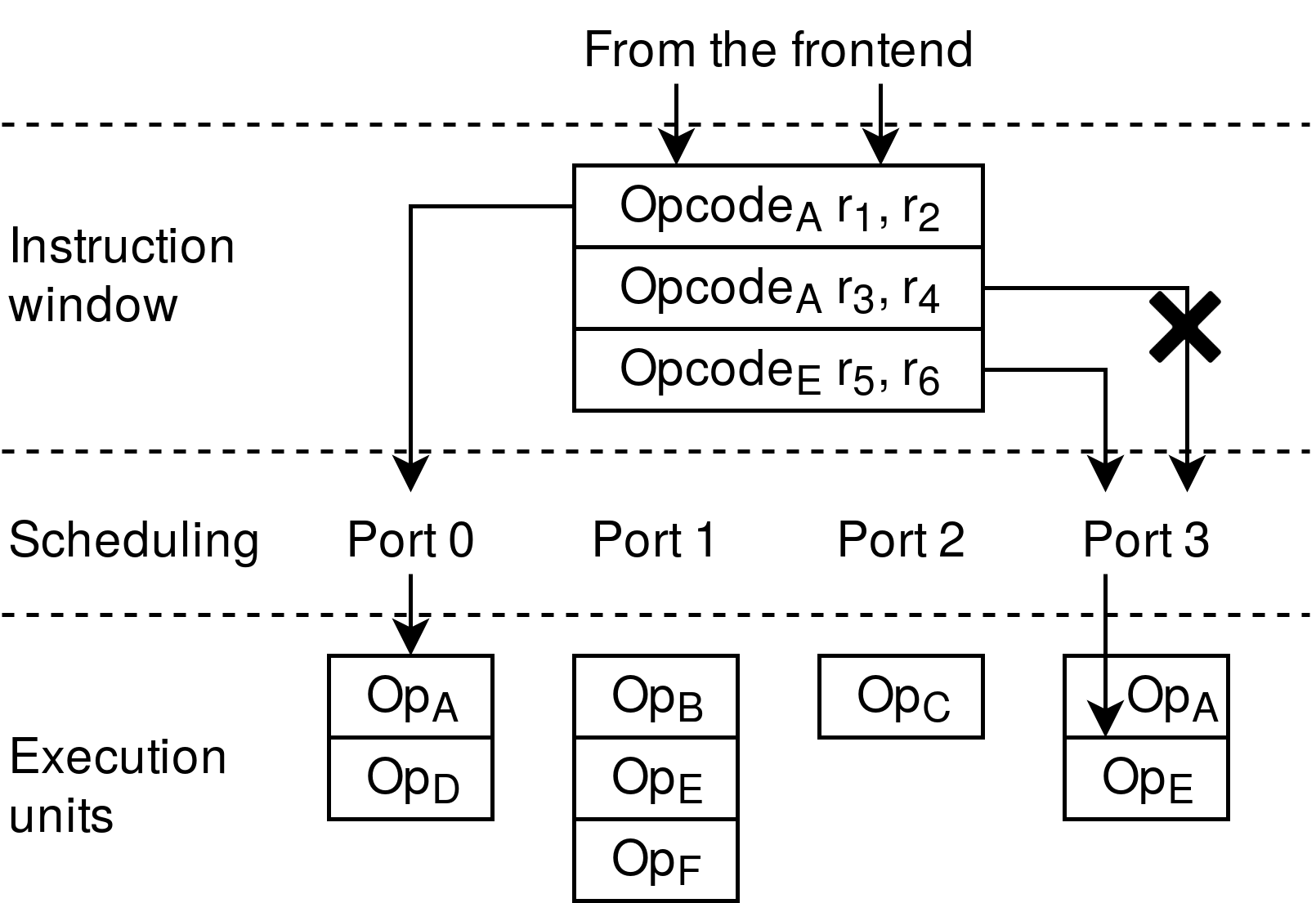}
      \caption{Instructions from the window are scheduled to ports shared by
               sets of execution units. A single instruction may be scheduled per
               port per cycle.}
    \label{fig:cpu_micro}
\end{figure}

\paragraph{CPU Microarchitecture} 
A modern CPU is typically split into two main components: the \emph{frontend}
and the \emph{backend} (or execution engine). The frontend predicts where to
fetch instructions from and creates a program-order stream of instructions to be
executed by the backend. The instructions are either decoded and executed
``as-is'' in RISC ISAs (e.g., IBM POWER or ARM) or broken down into RISC-like
instructions called $\mu$ops in CISC ISAs (e.g., x86 or IBM Z). For
brevity we refer to all instructions executed by the backend as $\mu$ops. Once
fetched and decoded, the $\mu$ops are placed in an \emph{instruction window}
(also referred to as issue queue or reservation stations) to be scheduled and
dispatched to execution units when their operands are ready. Every cycle, the
scheduler searches the instruction window to identify which $\mu$ops are ready
for execution and which execution unit is available to dispatch them to.
$\mu$ops can execute out of program order (e.g., a later $\mu$op in program
order can execute earlier) if their operands are ready and a relevant execution
unit is available. Ideally, all execution units would be designed to handle
every type of operation to maximize throughput. In practice, execution units are
specialized and only the more commonly used ones are replicated. A group of
execution units share a port, indicating their availability in a given cycle.
Contention for a port leads to delays in execution. \autoref{fig:cpu_micro}
demonstrates scheduling instructions from an execution window containing three
$\mu$ops, where contention for port 3 prevents the second $\mu$op from being
scheduled in the same cycle as the other two. 

\paragraph{Speculative Execution}
%
Because the stream of $\mu$ops is \emph{predicted} but is not guaranteed to
\emph{execute, complete and make its state visible to software}, the backend
also contains a \emph{re-order buffer} that commits the state of each completed
$\mu$op in program order to the software visible structures (i.e., register file
and memory). This execution of $\mu$ops is \emph{speculative} because the
frontend may have mispredicted the direction and/or the target address of a
branch operation. Upon misprediction, the pipeline flushes all $\mu$ops in the
re-order buffer and restarts fetching and decoding $\mu$ops. While executing on
the mispredicted path, the processor accesses the cache hierarchy leaving
side-effects which lead to cache-based side channels even though the values
accessed are discarded and do not impact the executing software.

\paragraph{Simultaneous Multithreading}
%
Out-of-order processors provision a large fraction of silicon area to mechanisms
that exploit speculation and parallelism in execution. While these mechanisms
are designed for peak parallelism, most structures (e.g., execution units,
branch tables, physical registers, instruction window, re-order buffer) remain
underutilized on average. Simultaneous MultiThreading (SMT) is a technique to
improve utilization of these structures by allowing $\mu$ops from multiple
threads (e.g., two in x86 and eight in IBM POWER) to execute simultaneously on a
single core. Individual SMT threads maintain their own architectural state, but
share many microarchitectural structures in the processor pipeline
simultaneously.  SMT (or HyperThreading as Intel brands its implementation) is
entirely transparent to software to which a single core appears as multiple
logical cores.  Besides the execution units, physical registers and instruction
window, it is an implementation's choice as to which other structures SMT
threads share.  Experiments have proven that the branch predictor can be shared
between hyperthreads~\cite{horn2018reading,coorporation2009intel} on Intel CPUs.

\paragraph{Speculative Execution Attacks}
%
Speculative execution can be exploited by priming the branch predictor with
sufficient history such that it is tricked into predicting the wrong target for
a branch. Because branch direction history (i.e., taken or not taken) 
is a shared resource, an attacking process can
prime the branch predictor of its victim. Similarly, a branch target buffer predicting 
the target address for a branch can be primed by an attacking process. This works for both
conditional branches as well as indirect branches.  In a conditional branch,
such as an array-size check in Spectre V1, the CPU can be tricked into
speculatively executing an out-of-bounds array access in spite of the failing
length check.  If the target address of the length check is not in the cache
then the memory fetch will take longer than the following speculatively executed
instructions.  In an indirect branch, the CPU can be tricked into speculatively
executing arbitrary code in a victim process by providing a malicious branch
history through a temporally or spatially (in the case of SMT) co-located
attacker process.
We discuss related work in~\autoref{sec:related}.

\paragraph{Cache-timing Side Channel}
%
\sloppy
Speculative execution attacks, such as Spectre, exploit the fact that a
speculatively executed and then discarded operation does have side effects on
the micro-architectural state, even if it has none on the  architectural state. 
For example, an instruction that operates on a value stored in memory will need to fetch
that value and cause the corresponding memory region to be pulled into the
cache.  The side-effect that the memory region is now cached is not undone when
the instruction is discarded instead of retired, and can be measured using cache
side channels.  For example, in Spectre V1 the victim code uses two dependent
array lookups, where the result of the lookup of the first array is used as an
index into the second array. This index can be leaked by measuring access times
to the second array through a flush and reload attack. By ensuring that the
second array has been flushed from the cache before the victim code
executes, and measuring the access times afterwards, only the lookup of the
index that has been used by the victim code will be significantly faster.

\section{Smother} 

In this section, we describe and evaluate \smother, a side channel based on 
port-contention, present in SMT architectures. \smother is based on the 
following observation: two co-located (i.e., running on the same physical 
core) hardware threads of execution share execution units. Instructions that are 
scheduled to execute on the same execution port will contend for the available 
resources. We show how this contention can be measured, at first in a 
coarse-grained way, i.e., with large sequences of instructions scheduled on the 
same port on both threads, and then in a fine-grained way, i.e., with minimal 
sequences of instructions. The result is that an unprivileged \emph{attacker 
process} can detect whether a co-located \emph{victim process} is running an 
instruction on a given port.

\subsection{Ideal covert channel}

In this experiment, we demonstrate port contention between two threads running
simultaneously on the same physical core and describe how it can be measured in
ideal conditions.

\subsubsection{Experiment design}
%
Executing instructions that occupy a specific port and measuring their timing 
enables inference about other instructions executing on the same port. We first 
choose two instructions, each scheduled on a single, distinct, execution port. 
One thread runs and times a long sequence of single $\mu$op instructions 
scheduled on port $a$, while simultaneously the other thread runs a long 
sequence of instructions scheduled on port $b$. We expect that, if $a=b$, 
contention occurs and the measured execution time is longer compared to the $a 
\ne b$ case.

\subsubsection{Experimental setup}
We run experiments on an Intel Core i7-6700K CPU running Ubuntu 16.04.4 stock 
kernel, version 4.15.0. Both attacker and victim are pinned to different 
hardware threads on the same physical core.  The CPU governor is set to 
\emph{Performance} for a constant clock frequency. The ``performance'' state is 
configured below the turbo frequency range to lower non-deterministic factors in
the environment. 
Apart from these changes, all other settings are 
kept to their defaults. Most notably, speculative-execution-related mitigations 
are left enabled.

In the measuring thread, we execute and time a sequence of 1,200 \Code{shl}, a
single $\mu$op instruction that executes on port 0 or port 6, which we denote
port 06, on this CPU. The colocated thread runs a sequence of either 1,200
\Code{shl} or \Code{popcnt} instructions: the \Code{shl} instructions directly
contend for port 06 while the \Code{popcnt} instructions will introduce no
contention as they execute only on port 1. Instruction-to-port mappings are
available through reverse engineering~\cite{fogInstTables} or the Intel
Architecture Code Analyzer (IACA) tool.

\subsubsection{Results and discussion}
%
We report averages over 10,000 runs, together with a 95\%-confidence interval 
calculated using the Student's t-distribution. The experiment successfully 
demonstrates that port contention occurs and that the \smother side channel can 
be used to extract information, as we can see in 
\autoref{tab:smother-port-contention}. Indeed, the run time of the contention 
experiment is about twice of the non-contended one. This indicates that port 
contention is likely the main bottleneck in this experiment.

This result shows how \smother can be used as a reliable covert communication 
channel between two co-located threads. However, as this experiment requires 
precisely choosing the type and number of instructions running in one of the two 
threads, it is yet unclear if port contention may serve as a practical side 
channel. We explore this aspect in the next section.

\begin{table}[h]
	\centering
	\begin{tabular}{lr}
          Experiment & Execution Time (cycles) \\ \hline
		Port contention &  1214 $\pm$ 67 \\
		No port contention & 674 $\pm$ 13 \\
	\end{tabular}
	\caption{Port contention covert channel: a thread running a long
        sequence of port 06 instructions is twice as slow when a co-located
      thread executes a long sequence of port 06 instructions, when compared
      to a co-located thread executing a long sequence of port-1-only instructions}
	\label{tab:smother-port-contention} 
\end{table}

\subsection{Characterization of the side channel}
\label{sec:smotherPoC}

We now analyse whether \smother is effective as a side channel for
distinguishing realistic sequences of instructions on a simultaneously
executing, co-located victim process. Specifically, we want to explore whether an
attacker can distinguish between the different sequences of instructions from a
known set which the victim may run. To encapsulate this property of the set, we
define the term \emph{\smother-differentiability}.

\paragraph{\smother differentiability}
%
Let us consider that the victim runs one sequence out of a set 
$V = \{V_0, V_1, ...\}$. The attacker is allowed to craft any sequence of
instructions $A$ and time multiple iterations of $A$ running concurrently with
the victim. If the attacker can infer which sequence $V_i \in V$ the victim was
running based on its timing measurements, the sequences in $V$ are said to be
\smother-differentiable. On its part, the attacker has a-priori knowledge of
what timing to expect when $A$ runs concurrently with each of $V_i \in V$. It
can use experiments in a similar, but controlled, environment to generate this
knowledge. Further, the attacker is allowed to use \emph{any} statistical test
or metric to make its decision. Examples of such metrics include the mean or the
median of the timings, or their distribution. 

In experiments in later sections of this paper, we shall establish
various pairs of sequences to be \smother-differentiable. After collecting
attacker timings distributions for each victim sequence in our controlled 
environment, we shall use the Student's t-test to establish statistical 
difference between them with at-least 95\%-confidence. 
We argue that an attacker, in an adversarial scenario, can correlate its 
own timing distribution with either of the a-priori distributions to 
identify the victim sequence.

At its core, \smother-differentiability implies that the sequences in $V$ have
differing degrees of utilization on some specific port(s) and vice-versa. The
attacker would ideally choose a sequence of instructions scheduled solely on
these ports to maximize the chance of encountering different levels of
contention across the different possible $V_i$. Through our experiments, we
wish to explore how short \smother-differentiable sequences can be and the ideal
length of attacker sequences to differentiate them. 

\begin{figure*}[!t]
\centering
\begin{subfigure}[b]{.48\textwidth}
        \centering
        \includegraphics[width=0.8\linewidth]{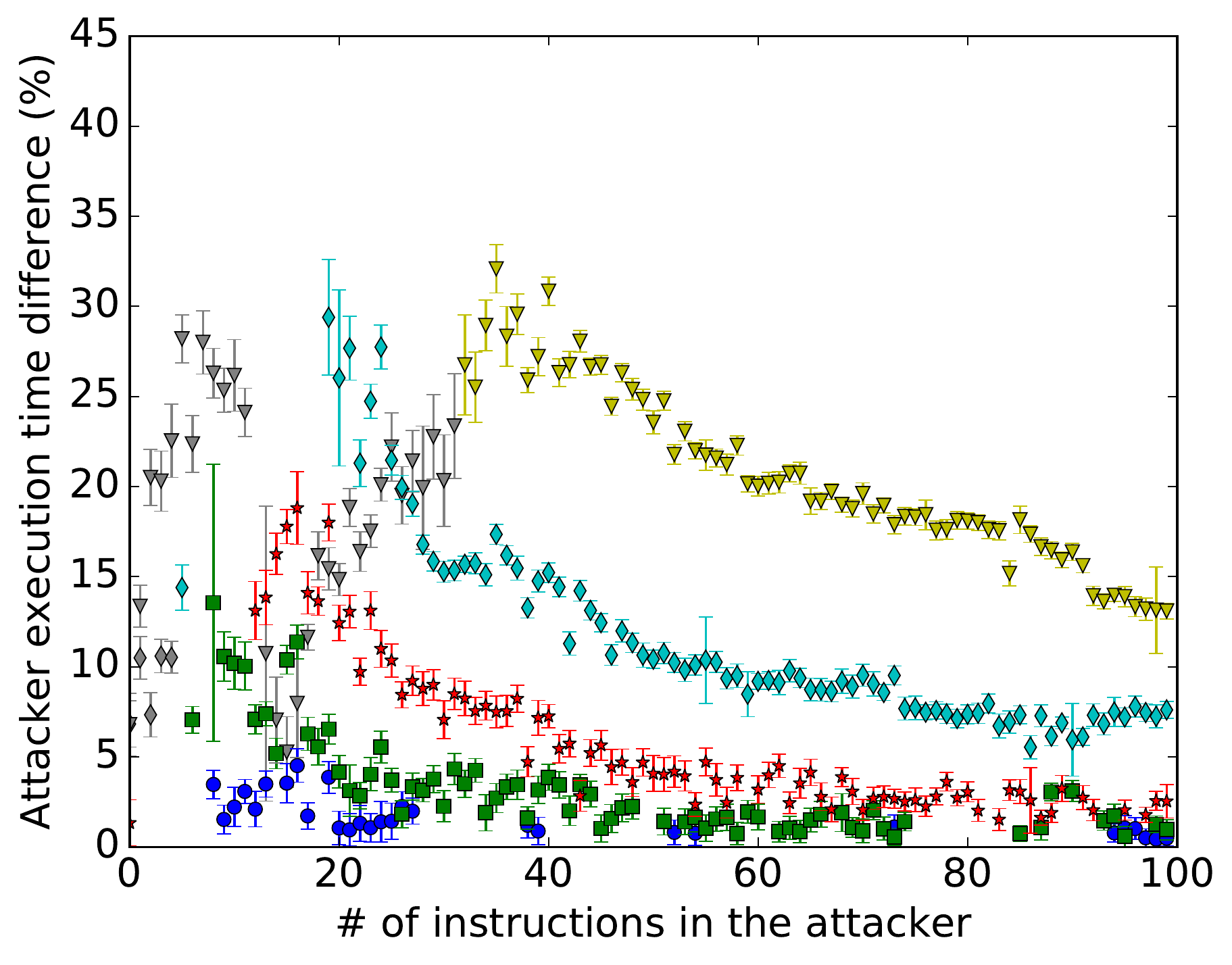}
        \caption{\smother attack using \Code{popcnt} to detect if the
        co-located victim runs on port 1.}
        \label{fig:smothercharac_popcnt_ror}
\end{subfigure}
\hfill
\begin{subfigure}[b]{.48\textwidth}
        \centering
        \includegraphics[width=0.8\linewidth]{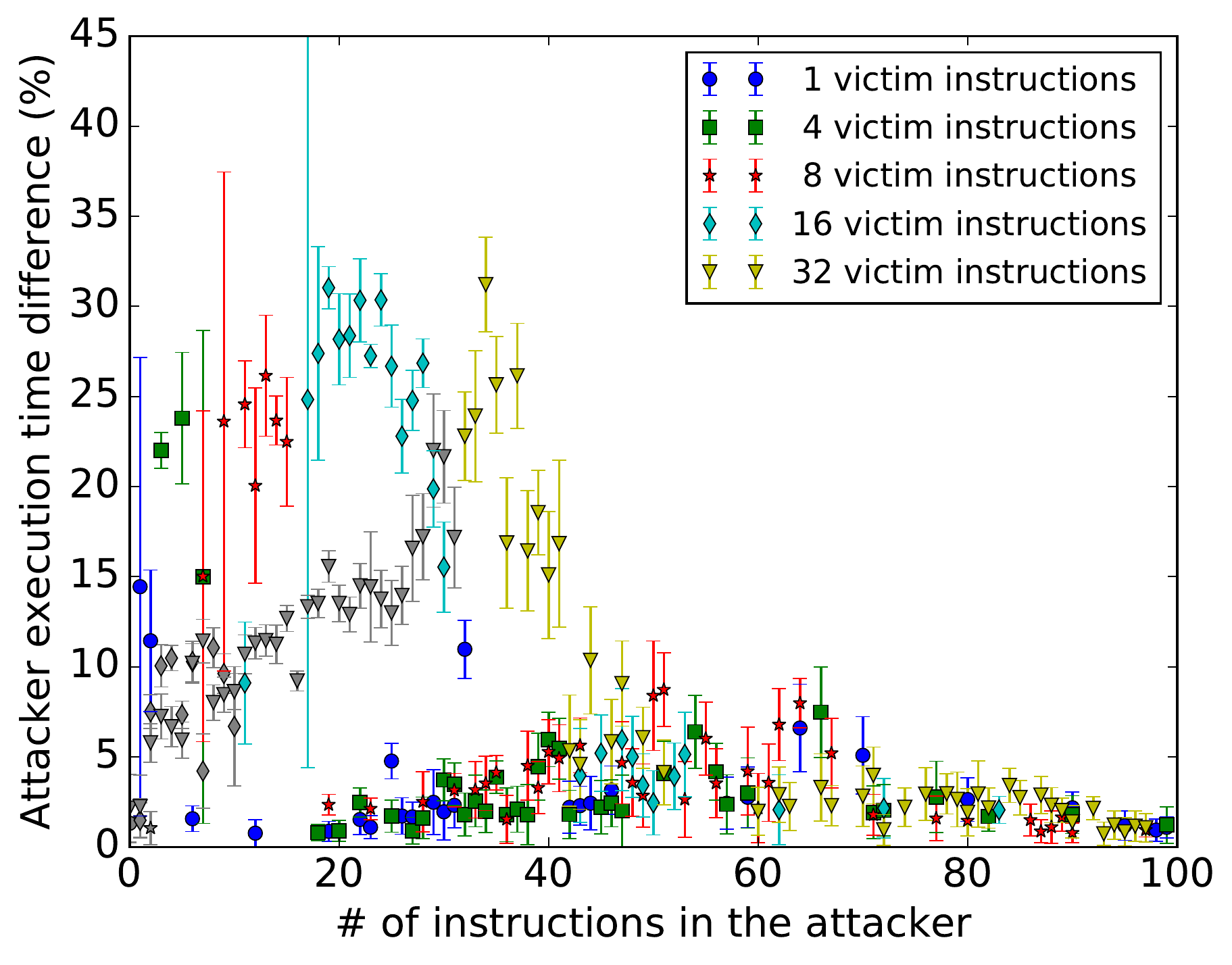}
        \caption{\smother attack using \Code{bts} to detect if the
  co-located victim runs on port 06.}
        \label{fig:smothercharac_multiport_cmovz_bts_popcnt}
\end{subfigure}
\caption{\smother side channel characterization. Each data point represents the
difference between the average execution time of the attacker thread,
between the port contention scenario and the baseline. We do not plot
the few data points where Student's t-test shows no statistically significant
difference between both distributions at 95\%-confidence. The
data points for which the attacker runs fewer instructions
than the victim are plotted in grey.} 
\label{fig:smothercharac}
\end{figure*}

\paragraph{Experiment design and setup}
In our first experiment, we consider a victim running sequences of either
\Code{popcnt}~(port 1) or \Code{ror}~(port 06) and an attacker timing a 
sequence of \Code{popcnt}.
We vary the length of both attacker and victim sequences, and check for
\smother-differentiability by noting the percentage change in mean execution time
for the attacker. In a second experiment, the victim runs either 
\Code{cmovz}~(port 06)
instructions or \Code{popcnt}. In this case, the attacker times a sequence of
\Code{bts}~(port 06) instructions with both operands as registers.

To run this experiment, an orchestrator process is used to fork the victim and
attacker processes, and to set their core affinities so that they share a
physical core.  We require the execution of the target sequence in the victim to
temporally overlap with the (timed) execution of the attacker sequence to assure
port contention. Therefore, the processes use a synchronization barrier which
ensures that any following instructions will be run concurrently. Thereafter,
each process runs their respective sequence, using \Code{rdtscp} to take
timestamps at the beginning and end of each run. The timestamps tell us the
number of cycles taken to execute the sequence and were used to also check that
the executions were properly synchronized. Atomic operations on variables in
shared memory were used to implement the synchronization. We repeat this process
to collect multiple timing samples.

In this set of experiments, we keep the same hardware and OS configuration as
used in the covert channel experiment, while precisely controlling the
synchronization of threads through the additional instrumentation described
above.

\paragraph{Results}
%
\autoref{fig:smothercharac} plots the average difference in attacker execution 
time between the two sequences of victim instructions for each experiment. The 
length of the sequence for the victim was taken from the set $\{1,4,8,16,32\}$
while the attacker sequence varied in length between one and 100 instructions. 

Our measurements confirm that timing short sequences of instructions is
feasible: for a vast majority of sequence-length combinations
the victim sequences were \smother-differentiable using the Student's
t-test on the attacker's running time distributions. While
timing \Code{popcnt}, 83\% of all combinations plotted in 
\autoref{fig:smothercharac_popcnt_ror} showed significant differences in
means between the victim's sequences of \Code{popcnt} and \Code{ror}. 

The measured differences vary from close to 0\% to ~40\%. Longer sequences of
instructions in the victim lead to higher differences and less variability in
measurements. Only 48\% of \Code{popcnt} measurements with sequence of 1 victim
instruction are \smother-differentiable, as opposed to 83\% for a sequence of 4,
and 100\% for a sequence of 32 victim instructions. This means that distinguishing a
sequence of one victim instruction (max. ~9\% difference and more variability) is
much harder than a sequence of 32 victim instructions (max. ~38\% difference and
less variability). 

We observe that there is an optimal number of attacker instructions to measure a
victim instruction sequence of a given length, which increases with the number
of victim instructions: from ~10 attacker instructions for one victim instruction
to ~45 instructions for 32 victim instructions. This is explained by the
following observations: contention for longer instruction sequences in the
attacker is easier to time, since attacker and victim sequences are more likely
to overlap. This effect fades when the attacker sequence becomes
significantly longer than the victim's, at which point only a small portion of
the executed instructions will contend, thereby leading to a smaller difference.

To show the breadth of possible \smother-differentiability results, we perform a
second experiment, with a victim running instructions which may be scheduled to
more than one port. Specifically, the victim runs either \Code{cmovz} (port 06)
or \Code{popcnt} (port 1). The attacker times a sequence of \Code{bts} instructions
(port 06) to measure the contention on ports zero and six.
\autoref{fig:smothercharac_multiport_cmovz_bts_popcnt} shows that multiport
instructions are still \smother-differentiable. However, variance is higher, and
we notice a steeper cut-off point beyond the optimal number of attacker
instructions. Indeed, intuitively, with more execution ports available, the
instructions are less likely to contend. In practice, this means the
attacker may need more runs to extract information, and the choice of the number
of attacker instructions is more important than in the previous experiment.
As in the previous experiment, we observe that the optimal number of
attacker instructions increases with the number of victim instructions. Beyond
this number, most experiments show lower \smother-differentiability, with most
between 0 and ~5\%.

While our results show that the \smother side channel exists and can be measured
even for a small sequence of instructions, we have noted a number of takeaways
and pitfalls to avoid during measurements, namely:
\begin{itemize}[topsep=0pt]
  \item Synchronisation of the target code sequence in the victim and the timed
code sequence in the attacker is extremely important, more so when the target
code sequence in the victim is short; 
  \item Pipeline bottlenecks other than port contention may occur and overshadow
the side channel. One such example is read-after-write hazards;
  \item The CPU may eliminate the execution of some instructions based on their
operands (one such case is \emph {zero idioms}). This results in those operands
not being executed, and removing contention;
  \item Some instructions (e.g., those from the SSE and AVX extensions) are
subject to aggressive power-saving features on modern CPUs. This makes measuring
port contention more difficult (and the power savings may in fact serve as its
own side channel~\cite{schwarz18arxiv} separately from \smother).
\end{itemize}

Finally, we note that practical instruction sequences are unlikely to be
identical repeated instructions. However, this is not required for practical
\smother side channels: it is only required that, among a sequence of
instructions, they exercise different degees of port pressure on the port 
that the attacker
is measuring. We further expand on this idea in \autoref{sec:discovery} for
practical \smother-differentiable sequences.

\section{\smotherspectre} 
\label{sec:smotherspectre}

\smotherspectre is a speculative code-reuse attack technique which starts at an 
indirect jump on the victim's usual execution path. The attacker leverages 
Branch Target Injection (BTI) to ``poison'' the CPU's branch predictor such that 
when the victim's fetch unit asks for the target of the indirect jump, it is sent 
the address of a separate data-dependent conditional jump within the victim's 
binary with \smother-differentiable fall-through and target sequences. During
the period of the speculative execution, the victim evaluates the 
condition and jumps to either the target or fall-through sequences. The 
attacker times a sequence of relevant instructions to identify which sequence is
run on the victim (\smother), thereby inferring the outcome of the 
condition and learning some information about the victim's data.

\smotherspectre complements and
extends existing attacks~\cite{kocher18oakland, koruyeh18woot, canella18arxiv}
which use cache-based side channels to exfiltrate secrets. Using such channels
implies that these exploits
\begin{inparaenum}[\itshape i\upshape)]
  \item require the presence of \emph{special} gadgets in the victim code, or
the ability to inject them; and
  \item depend on speculative execution leaving persistent, measurable
microarchitectural side-effects.
\end{inparaenum}

Calls using function pointers in C/C++ are traditionally implemented by indirect
calls in assembly. While exploitable indirect jumps are prevalent in most
programs, the first observation limits the set of available gadgets for
ultimately leaking secrets. This scarcity, along with the overheads of some
software-only mitigations, justifies the use of user-space programs to not
deploy countermeasures such as retpolines or STIBP by default.
In contrast, \smother-differentiable gadgets are easily found (as we demonstrate
in \autoref{sec:discovery}). Almost every conditional jump can be part of a 
\smother-gadget, requiring only its fall-through and target to be 
\smother-differentiable. For example, \Code{libcrypto} from the OpenSSL library
contains more than 12,000 readily usable gadgets.

The second observation has lead to the proposal of defenses that ensure that
\textit{all} changes to microarchitectural state be
undone~\cite{khasawneh2018safespec}. However, the port-contention based
side-channel persists even if the CPU were able to perform
a perfect roll-back of changes caused by non-retired instructions.  The very
fact that instructions are speculatively executed remains a measurable quantity.
These characteristics allow \smotherspectre to present a more powerful avenue of
attack.

In this section, we first present the attacker model and objectives for \smotherspectre.
We then explain the basic premise of the attack, the conditions required and how
we ensure these are met in our proof-of-concept. We then present a
characterization of the \smotherspectre side channel. Finally, we discuss the 
characteristics of some \smother-gadgets we found in common system libraries, and
what information they may be used to leak.

\begin{figure}[]
    \centering
    \includegraphics[width=0.9\linewidth]{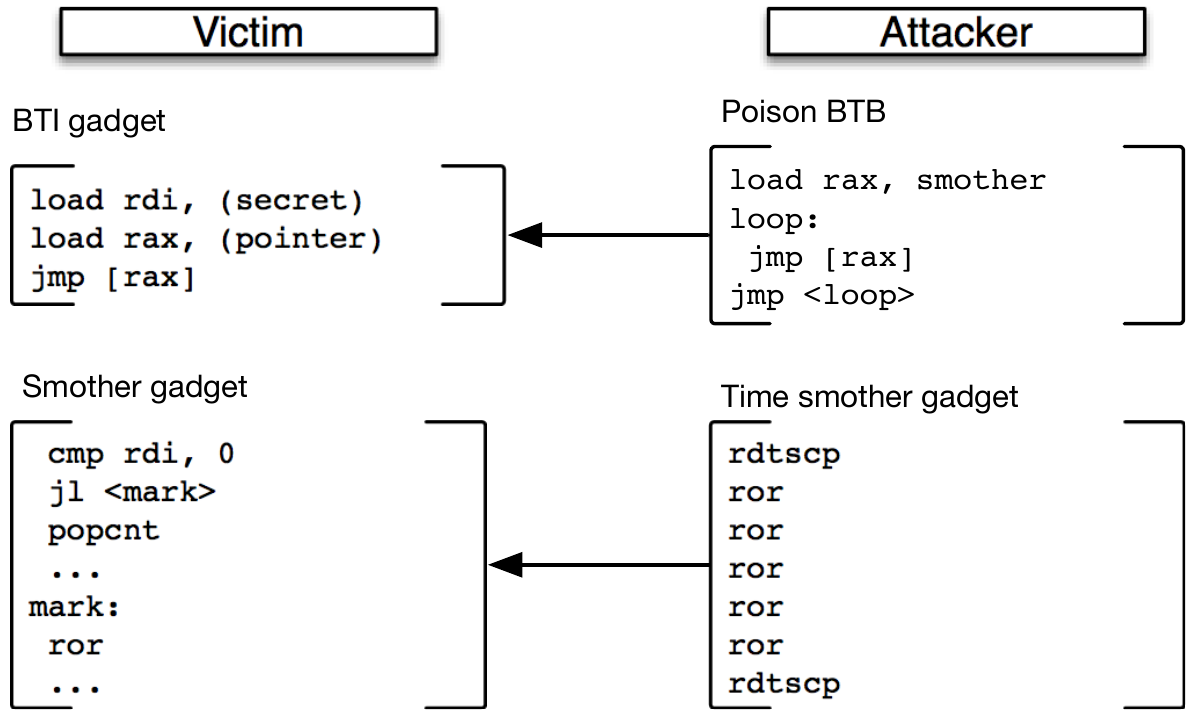}
    \caption{Overview of the \smotherspectre components.}
    \label{fig:smst}
\end{figure}

\subsection{Attacker model}

The objective of a \smotherspectre adversary is to extract secret information from 
a victim process and we make the following assumptions about the attacker: 
\begin{inparaenum}[\itshape i\upshape)]
  \item they control code in a process co-located with the victim process; 
  \item they can launch branch target injection attacks.
\end{inparaenum}

The first assumption is justified: if the attacker can execute code on the same 
machine of the victim, the scheduler may schedule the attacker and victim on two 
different threads of the same physical core. An example of such colocation may
exist in public cloud offerings where compute resources are shared at a fine 
granularity between tenants: for IaaS, virtual cores for different customers 
may map to the same physical core, for PaaS/SaaS processes for different 
tenants may be similarly scheduled~\cite{ccccloudsuite, gcloudfaq}. 

Existing mitigations against BTI include software (retpolines) and a set of 
hardware interfaces for flushing the indirect branch predictors at the 
appropriate times and for not sharing them across SMT threads 
(IBRS, IBPB and STIBP in Intel).
These mitigations come with a potentially severe performance
impact~\cite{linusemails}. As such, these controls have been enabled only for
selected system components such as the kernel, and none of the user-space
programs we have analysed make use of them. 
The adversary also needs to know the victim's code base, which is possible
through the use of common libraries and open-sourced applications, 
and where it is located in memory. It must be able to circumvent ASLR and similar
controls: the literature contains several examples~\cite{evtyushkin2016jump,
sotirov2009bypassing, hund2013practical} of how this is achievable in practice, 
including an approach using the same BTB weaknesses that make BTI possible.

\subsection{Attack principle}
\label{sec:attack_principle}

\autoref{fig:smst} shows a side-by-side layout of the code of a 
victim and an attacker in the \smotherspectre setting. As the figure shows, the 
attack requires two types of gadgets in the victim code:
\begin{itemize}
  \item A BTI gadget: Stores secret data into memory or a register (called the 
\textit{\smotherspectre target}) followed by an indirect branch that can be 
poisoned by the attacker;
  \item A \smother gadget: A data-dependent conditional jump whose control variable 
  is the \smotherspectre target, with \smother-differentiable (see 
  \autoref{sec:smotherPoC}) target and fall-through code paths.
\end{itemize}
The example BTI gadget in \autoref{fig:smst} stores the secret into 
the register \Code{rdi}, a pointer into \Code{rax} and finally jumps 
to the location pointed to by \Code{rax}. The corresponding \smother gadget
contains an \Code{rdi}-dependent conditional branch where the jump target and
fall-through contain \smother-differentiable instruction sequences (\Code{popcnt} 
and \Code{ror}). 

Note an important difference between traditional data-dependent control flow
sequences and \smotherspectre. Data-dependent control flow sequences over
confidential data are considered vulnerabilities, especially when found in
cryptographic libraries. \smotherspectre does not require such a vulnerability to
be present in the victim. It connects the loading of a secret variable to a
register or memory location (BTI gadget) with an altogether independent,
speculatively executed sequence, which happens to perform a compare-and-jump over
that same register or memory location (\smother gadget). The two sets of instructions
may well be entirely uncorrelated from a software development perspective,
making the pattern harder, if not entirely impossible, to eliminate.

The attacker proceeds in two main steps, as shown in 
\autoref{fig:smst}: in the first phase the attacker 
performs traditional, Spectre v2 style BTI and then enters 
in a busy wait sequence, for instance a sequence of \Code{nop} instructions. 
The purpose of the latter is to align the second phase of the attack with the 
speculative execution of the \Code{mark} or \Code{fall-through} sequence in 
the victim. In the second phase the attacker performs a \smother-style timing of 
a carefully selected sequence of instructions --  \Code{ror} in the example. 
The attacker then proceeds to a statistical analysis of the gathered timing 
information to learn one bit of information. This entire process can be repeated 
with different gadgets to leak different bits, and thereby reconstruct the 
secret. Note that while the example utilizes the indirect-branch prediction 
hardware to steer speculative execution to gadgets, any existing branch 
redirection method may be used for this purpose (for example the return stack buffer).

\subsection{Characterization of the Side Channel}
\label{sec:smotherspectrepoc}

In our experimental testbed to characterize the \smotherspectre side channel, an 
orchestrator process forks a victim and an attacker 
process, pins them to two threads on the same physical core and executes an 
attacker and a victim process (similar to the testbed in \autoref{sec:smotherPoC}). 
Attacker and victim processes execute the body of a
loop after synchronization using atomic operations on shared memory. The body of
the loop is constructed as described in \autoref{fig:smst}.

In our proof-of-concept, we leverage the branch target buffer to redirect an
indirect branch in the BTI gadget of the victim to the \smother
gadget. In order to maximize the success rate, we
\begin{inparaenum}[\itshape i\upshape)]
  \item insert a series of N always-taken branches just prior to the indirect
branch;
  \item ensure that the addresses of the branches (including the target of BTI)
are located at congruent addresses between attacker and victim;
  \item disable ASLR;
  \item evict the cache-line containing the indirect jump pointer.
\end{inparaenum}
As other works have shown, the random ASLR offset can be leaked in a 
real-life attack~\cite{evtyushkin2016jump,sotirov2009bypassing}, and 
BTI can be performed by aliasing addresses
(in the BTB) with very high success rates~\cite{horn2018reading}. Therefore, we
disregard these factors while creating our proof-of-concept (PoC). Evicting the
jump pointer allows us to extend the duration of the victim's speculative 
execution, in order to establish an upper
bound for accuracy and throughput for the channel. In alternate settings, we 
have noticed that usual victim computation can evict the pointer from the L1 cache. 
The resulting period of speculation is enough for our attack to work. 

\sloppy
Further, we introduce instrumentation to obtain information about the success of 
the BTI attack. This information is supplied by the Performance Counter Monitor 
(PMC) infrastructure and can be obtained by using the \Code{msr} kernel module. We 
use it to program the PMC counters to retrieve samples for the 
\Code{BR_MISP_EXEC.TAKEN_INDIRECT_JUMP_NON_CALL_RET} event, which is triggered 
every time the target of a taken indirect jump is mispredicted. PMC counters are 
sampled at the start of every loop and once more at their end. BTI is successful 
whenever the difference in the value of the counter is 1, given that the victim code
contains only one indirect jump.

The timed instruction sequence in the attacker consists of a series of 42 
\Code{crc32} instructions operating over randomly chosen, nonzero values. 
The victim process contains an equivalent sequence of \Code{crc32} 
instructions at the fall-through of the branch: given that \Code{crc32} 
instructions execute exclusively on port 1, if BTI is successful and the 
speculated conditional branch is not taken, the victim will be competing for 
execution on port 1 with the attacker. The target of the branch instead contains 
a sequence of instruction designed to be executable on more ports (0,1,5,6) and 
thus display less contention with the attacker.

\begin{figure}
    \includegraphics[width=\columnwidth]{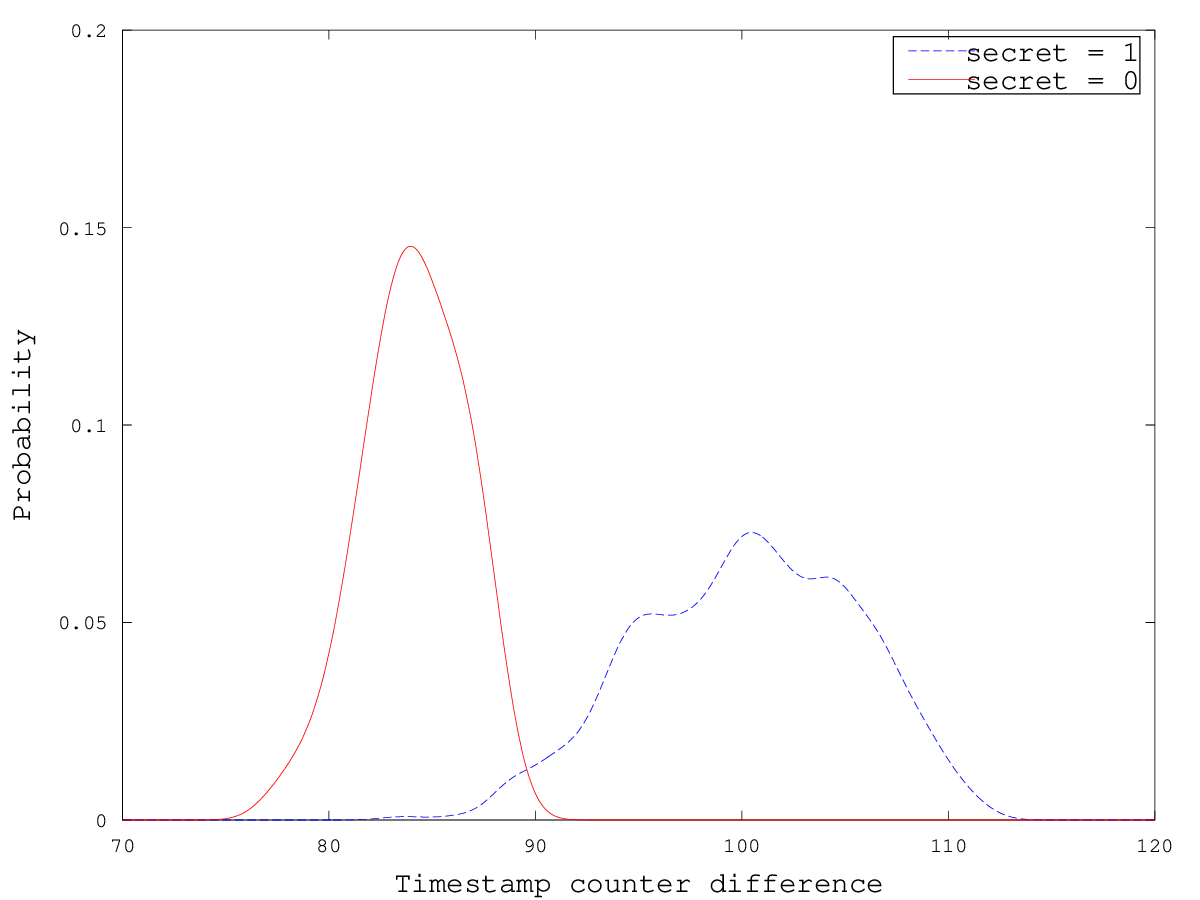}
    \caption{Probability density function (estimated using kernel 
density estimation) for the timing information of an 
attacker measuring \Code{crc32} operations when running concurrently with a 
victim speculatively executing a \smother-gadget. }
    \label{fig:smotherspectre_distributions}
\end{figure}

We collect two sets of samples: one when the victim's secret is set to 
zero, and one where it is set to a nonzero value. 
\autoref{fig:smotherspectre_distributions} shows the results of the experiment on a 
Skylake platform (i7-6700K). As we can 
see, the distributions obtained when the victim has a non-zero secret generates 
more contention on port 1 and thus causes the attacker to measure a higher 
time-stamp counter difference. This is justified by the fact that a nonzero 
secret causes speculative execution to be directed to the fall-through of the 
branch, which we have designed with a competing sequence \Code{crc32} 
instructions.

In the next phase of the attack, we use the results of this experiment as 
profiling information to read the side channel. To this end, a bit sequence is 
generated and set - bit by bit - as the secret value on the victim. 
Based on the results of \autoref{fig:smotherspectre_distributions} we choose a 
time-stamp counter difference of 89 as a threshold: if the mean of the samples 
is higher than the threshold we conclude that the secret is 1, and 0 otherwise.
The experiment is run, 5 samples each are collected for $20,000$ secret bits. 
The attacker is able to correctly guess the victim's secret with a success rate
of over $98\%$. The entire experiment takes $0.83s$ as reported by \Code{time}
yeilding a sample rate of 120,000 samples/second, and a leakage rate of 
$24,000 bit/s$. A similar experiment with guesses based on 1, 2, 3 and 4 samples
result in accuracies of $72\%$, $78\%$, $83\%$, and $90\%$ respectively. 
As expected, there is a trade-off between accuracy and leakage rate.

We repeated this experiment on a Haswell processor (i7-4790) using the same
attacker-timed sequence and victim's \smother-gadget. With a threshold of 85 cycles,
the attacker was able to guess the victim's secret bit with an accuracy of $53\%$, $62\%$,
$69\%$, $70\%$ and $76\%$  based on 1, 2, 3, 4 and 5 samples respectively. We also 
validated the attack on a Broadwell processor (i3-5005u).

\subsection{Discussion about \smother-gadgets}

\begin{table}
\centering  
{\sffamily\fontsize{8}{10}\selectfont
\begin{tabulary}{\linewidth}{Lcccccccc}
\toprule
    &    glibc  & ssl & pthread & ld &  crypto & z &  stdc++ & Together  \\
\midrule
rax &    14 & 12  & 9       & 7  &  11     & 8 &  8      & 21   \\
rbx &    6  & 2   & 0       & 1  &  6      & 1 &  1      & 9    \\
rcx &    8  & 1   & 2       & 1  &  5      & 1 &  2      & 8    \\
rdx &    10 & 2   & 5       & 6  &  7      & 2 &  3      & 14   \\
rsi &    8  & 4   & 1       & 2  &  3      & 1 &  1      & 10   \\
rdi &    8  & 2   & 0       & 2  &  3      & 1 &  0      & 11   \\
rsp &    2  & 0   & 0       & 0  &  0      & 0 &  2      & 3    \\
rbp &    5  & 3   & 0       & 0  &  9      & 0 &  0      & 13   \\
\bottomrule
\end{tabulary}
}
  \caption{Number of different leakable register bits (out of 64) using \smother 
  gadgets from common system libraries, specifically \Code{test-jxx} 
  one at a time, on multiple iterations of the victim with the same register state.}
  \label{tab:bits_leaked}
\end{table}

A \smother gadget is defined by two attributes: the condition on which the jump depends and 
the sequences on both paths following the jump. The latter determines whether the gadget can 
be used in a \smotherspectre attack: the sequences must be \smother-differentiable. The former 
determines the information leaked by the gadget. In this section, we shall discuss some of 
the \smother-gadgets found in real libraries, and what they can leak.

Common instructions which set the condition flags in \smother-gadgets we found 
are \texttt{cmp}, \texttt{test}, \texttt{add}, \texttt{sub}. \texttt{cmp-jxx} 
sequences compare a value in a register (or loaded from memory) against other 
registers or against a constant. Each gadget reveals a constraint on the value. 
\texttt{test-jxx} and \texttt{and-jxx} sequences perform a bitwise-and of two 
values, setting flags based on the result. When one of the values is a constant 
with a single bit set, the gadget can be used to test whether specific bits 
are set in the first operand. Such gadgets reveal the corresponding 
bits to the attacker. When the second operand is not a constant, but a register 
whose value may be predicted or controlled, the attacker gains the power to check 
bits other than those specified by the constant gadgets.

Of the over 12,000 gadgets found in \texttt{libcrypto}, approximately:
\begin{itemize}[noitemsep, topsep=0pt]
\item 2,800 are \texttt{cmp} operations 
\item 3,900 are \texttt{test} operations
\item 1,500 are \texttt{add} operations
\item 970 are \texttt{or} operations
\end{itemize}

There are around 350~\texttt{cmp-jump} gadgets which compare a value in a 
register or in memory against zero. Around a 100 gadgets, which check for 
greater-than/lesser-than conditions against zero, can be 
used to leak whether the value is positive or negative. Another 294~gadgets 
compare against (the constant) one, and 807 gadgets compare against other constants. 
Around 370 \texttt{cmp-jump} gadgets have a memory operand of which 118 compare with 
non-zero constants. 300+ gadgets compare with values on the stack, of which 33 are 
against non-zero constants. Of all \texttt{cmp-jump} gadgets, more than 400 check for 
signed or unsigned greater-than. The number of signed or unsigned lesser-than is 
about the same. 

For victims running in a loop, there are cases where the register/memory state of interest 
will be the same across iterations. For example, a register/memory location storing a 
secret, cryptographic key can be expected to hold the same value across multiple calls to 
the encryption function. The attacker can leverage the BTI gadget to redirect the victim to 
different \smother-gadgets on different iterations, each time leaking different information 
about the secret. Over multiple iterations, the attacker can effectively leak multiple bits 
of the same secret, chaining the leaks from different \smother-gadgets. In 
\autoref{tab:bits_leaked}, we show how many bits we can leak from the registers by chaining 
\smother-gadgets found in commonly used system libraries. To illustrate this for
one specific register, \autoref{app:rax_gadgets}
lists the gadgets which can be chained to leak 21 bits from \Code{rax}.

\section{Gadget discovery}
\label{sec:discovery}

As described in~\autoref{sec:attack_principle}, we require two gadgets to be
present in the victim code for \smotherspectre.  We investigate the characteristics
of ideal gadgets and how to find them in a given piece of code.  We introduce
\textit{port fingerprinting} to summarize the port utilization of an instruction
sequence and assess the potential to be detected using \smother.  Port
fingerprinting enables a comparison of the port utilization of two instruction
sequences and rank combinations of instruction sequences based on their
difference in port utilization.

\paragraph{BTI Gadget}
%
The purpose of the BTI gadget is to pass the secret through a register to an 
arbitrary code target in the same process.  Depending on the attack scenario, the 
BTI gadget is the only piece of code that is strictly required to be present in 
the victim.  Ideally, it just consists of two instructions: one that moves the 
secret into a register and an indirect control-flow transfer.  In order to 
maximize the speculative execution window, the target of the indirect 
control-flow transfer should be retrieved from uncached memory.  An archetype of 
an ideal BTI gadget is a virtual function call in C++, with the secret value 
being an argument to such a function call. In the System V x86\_64 calling convention, 
the first six parameters of a function are passed in registers. Further, the 
typical implementation of a virtual function call uses indirection through a 
vtable to resolve the binding at runtime.  Since the vtable is stored in memory, 
the target of the call needs to be loaded, which can cause a speculation window of 
upto a few hundred (\textasciitilde200) cycles if the vtable has been evicted from the cache 
prior to the 
call. We can reasonably assume that this will happen in practice if 
objects are created by an early initialisation phase and used (potentially much) 
later in response to external events. Similarly, calls to functions in
dynamically-loaded ELF (Executable and Linkable Format) libraries
also employ an indirect jump, using a 
pointer from the Global Offset Table (GOT) to facilitate dynamic symbol 
resolution. Arguments in such calls may contain sensitive information which can
be compromised by an attacker using these jumps as BTI gadgets.

\paragraph{\smother Gadget}
%
A \smother gadget is the receiving end of a BTI gadget.  Depending on the
attack scenario, it is either already part of the victim, or can be supplied
via an additional attack vector.  It starts with an instruction that compares
the register to a known value.  The known value can either be a known
immediate in the code, or, more powerfully, an attacker-controlled value
specified via an additional attack vector.  The next instruction needs to be a
conditional control flow transfer based on this comparison leading to 
\smother-differentiable sequences.  To maximize the
chances of \smother-differentiability, the instruction sequences should each have a
distinct port fingerprint such that they can be clearly distinguished from one
another. This depends on the layout of the execution engine: on
Intel Skylake, a prime example would be one branch with a sequence of
\Code{AES}  instructions (only port 0) and another branch with
a sequence of \Code{MMX} instructions, predominantly limited to port five.
Besides, the instructions should ideally not load from or store to memory, as
potential cache misses introduce noise. Further, the more generic the
instructions in the sequence are, the more likely it is that their execution
unit does not require a warm-up phase during which execution is slow, again
introducing noise.

\subsection{Ranking \smother-gadgets} 

The instruction sequences we consider consist of basic blocks that start at the 
respective branch targets. To identify instruction sequences that are ideal for 
\smother and compare them against one another, we need to measure their 
suitability for \smother. The primary criterion is that the compare instruction 
operand has to match the register that is loaded with the secret in the BTI 
gadget. Further, we evaluate the instruction sequence at the branch target and 
fall-through by quantifying three properties:
\begin{inparaenum}[\itshape i\upshape)]
\item the port utilization difference of the two branch targets ($r_p$),   
\item the difference of the two branch targets in terms of the length of the
branches ($r_l$), and   
\item the amount of memory operations in both branches ($r_m$).
\end{inparaenum}
To compare instruction sequences based on these properties, we combine them
using the rank product $RP(g)=(\prod_{i=1}^{k}r_{gi})^{1/k}$ for our $k (=3)$
properties.

To compare the port utilization, we first use Intel's Architecture Code
Analyzer (IACA) to obtain a port fingerprint $P$ for a given instruction
sequence. The port fingerprint is a summary that lists the total number of
cycles spent on every port for a given instruction sequence $P={p_0\dots
p_7}$. IACA internally uses a microarchitecture-specific model of the
processor to compute the cycles, taking out-of-order execution into account.
It also models the divider pipe on Skylake, allowing port zero, which handles
the complex \Code{div} instruction, to be ready for the next $\mu$op in the
next cycle, while the \Code{div} is still being executed. As it cannot know
better, IACA assumes all CPU resources to be fully available prior to
execution of the sequence. An open-source alternative to IACA,
OSACA~\cite{laukemann18osaca} also supports AMD processors.

To compare two port fingerprints $P$ and $Q$, we subtract them and
then calculate the utilization difference as the sum over the
vector: $r_p=\sum^{i}_{0..7}(|p_i-q_i|)$.  The larger $r_p$, the higher the
difference in port utilization of the two instruction sequences.  The
utilization difference will be high for long instruction sequences that do
not share a port.  Such instruction sequences lend themselves well to
\smother.

While a ranking based on the port utilization difference already captures the
most important aspect, it has one drawback: gadgets where the branch
instruction sequences are of different length, such as 2 instructions vs. 20
will rank high, whereas we prefer sequences of equal length for the timing.
Therefore, we also include the inverse of the length difference
$r_l=abs(l_1-l_2)$ between the sequences of a gadget in the ranking.

Finally, we also take the potential noise into account that can be caused by
memory operations. On our targeted Skylake processors, the ports 2, 3, 4 and 7
are used for scheduling these. We include the inverse of the sum, 
$r_m$, of the cycles spent on these ports in both branches as an additional 
ranking for the
gadget. The final rank of a gadget $g_i$ is given by ${RP(g_i)=(r_{p_{i}} \cdot
(max(r_l)-r_{l_{i}}) \cdot (max(r_m)-r_{m_{i}}))^{1/3}}$.

\subsection{Finding Gadgets} 
We develop a tool
to aid gadget discovery based on the popular distorm3 disassembler and Intel's
Architecture Code Analyzer, and use it to analyze a number of common system
libraries that are likely to be linked to a victim executable. For the
analysis we only consider gadgets with a branch length between 3 and 70
instructions, with 3 instructions being a reasonably low bound for smothering
and 70 instructions being an upper bound for speculative execution. 
Our search looks for valid instruction sequences starting at every offset in the
binary. Therefore, it would detect any \smother-gadget resulting from an 
unintended sequences of bytes (starting from the middle of an intended instruction)
which might decode to valid instructions.
We show
the results in~\autoref{tab:gadgets}, the libraries analyzed are taken from a
regular Ubuntu 18.04 LTS installation. We focus on \smother-gadgets that
compare against the registers used in the x86\_64 calling convention and
either use the value in the register directly, or use it as a pointer and
compare to a value pointed to in memory. The rationale behind this is that BTI
gadgets are typically indirect calls that pass a secret, such as a
cryptographic key, as a parameter. The results show that we can find enough
\smother-gadgets even in a single common library such as \Code{glibc} alone. Note
that this method applies irrespective of whether the library is loaded at
runtime or is statically linked into the victim's binary. However, none of the gadgets
found were formed from instructions decoded from unintended byte sequences.

\begin{table}
  \centering
  {\sffamily\fontsize{8}{10}\selectfont
  \begin{tabulary}{\linewidth}{Lcccccccccccc}
  \toprule
                   & RDI  & RSI   & RDX   & RCX   & R8 & R9  \\
  \midrule
  glibc 2.23       & 1155 & 1502 & 3864 & 4256 & 568  & 615 \\
                   &1040 & 932 & 257 & 1029  & 135 & 29 \\
  stdc++ 6.0    & 189 & 400 & 869 & 1399 & 97 & 73 \\
                   & 209 & 65  & 98  & 276  & 58 & 14 \\
  ld 2.23          & 105 & 130 & 412 & 359 & 41 & 31 \\
                   & 46  & 47  & 29  & 110 & 6  & 0 \\
  pthread 2.23     & 23 & 56 & 70 & 82 & 25 & 8 \\
                   & 23 & 2  & 7  & 34 & 3  & 0 \\
  z 1.2.11      & 76 & 85 & 138 & 338 & 66 & 80 \\
                   & 24 & 29 & 8   & 96  & 16 & 5 \\ 
  crypto 1.1    & 1132 & 1048 & 1659  & 2566 & 45 & 29 \\
                   & 310 & 319 & 224 & 1036 & 239 & 167 \\
  ssl 1.1       & 243 & 239 & 376 & 500 & 39 & 21\\
                   & 95  & 32  & 29  & 239 & 12 & 1 \\
  \bottomrule
  \end{tabulary}
  }
  \caption{\smother-gadgets we found in common system libraries, for the
registers used to pass arguments in the System V x86\_64 calling convention. First line: number of \smother-gadgets that use the value in the register, second line: number of gadgets that use its pointee.}
  \label{tab:gadgets}
\end{table}

One under-approximating limitation of our gadget search algorithm is that it assumes that 
gadgets start from the latest 
flag-setting instruction before the jump. Suppose a sequence in the victim's code is 
\Code{shl 8, rax; test 1, rax; jz 0xadd;}. Our tool will find 
\Code{test 1, rax; jz 0xadd;} as a 
\smother gadget which leaks the least significant bit (LSB) of \Code{rax}. However, the 
instructions preceding this might perform computations which cause the 
gadget to leak different information. The 
entire sequence is a different \smother gadget which leaks the 9-th least significant bit of 
\Code{rax}. The space of usable \smother-gadgets exceeds the ones
we have found, and depend on the particular victim's code.

\section{Real world attack} 
\label{sec:realworldattack}

We demonstrate real-world attacks on OpenSSH and OpenSSL, two commonly used 
programs handling sensitive secrets that have been extensively hardened against 
regular and side-channel attacks.

\subsection{OpenSSH attack}
\label{sec:opensshattack}

\begin{figure*}[!t]
    \centering
    \begin{subfigure}[b]{.2\textwidth}
      \begin{minilisting}
.rept 8;
addl r8d, r9d;
addl r10d, r11d;
addl r8d, r9d;
addl r10d, r11d;
.endr;
      \end{minilisting}
      \caption{Attacker-timed code}
      \label{fig:real_attacker-smother_ssh}
    \end{subfigure}
    \begin{subfigure}[b]{.30\textwidth}
      \begin{minilisting}
0x6f8dc:  testl 0x100, (rdi)
0x6f8e2:  je     6f8ef
0x6f8e4:  mov    0x10(rbx),rax   
0x6f8e8:  sub    0x8(rbx),rax
0x6f8ec:  sub    rax,rsi
0x6f8ef:  mov    rbx,rdi
...
      \end{minilisting}
      \caption{Victim {\smother} gadget (\texttt{glibc})}
      \label{fig:real_victim_smother_ssh}
    \end{subfigure}
    \begin{subfigure}[b]{.4\textwidth}
      \begin{minilisting}
static void (* volatile ssh_bzero)
                (void *, size_t) = bzero;
void explicit_bzero(void *p, size_t n) { 
  ...
  ssh_bzero(p, n);
  ...
}
    \end{minilisting}
    \caption{Victim BTI gadget (OpenSSH)}
    \label{fig:real_bti_gadget_ssh}
  \end{subfigure}
  \caption{Gadgets from real-world libraries used
          in our {\smotherspectre} exploit for leaking the $7{th}$ least 
      significant bit of \Code{rdx}'s pointee}
  \label{fig:real_smother_gadget_ssh}
\end{figure*}

OpenSSH is widely used to securely and privately connect to servers over 
untrusted networks. The confidentiality of the server's private key is essential 
to the security of the overall system. Leaking the private server key allows an 
attacker to impersonate the server, acting as a man in the middle. In the 
OpenSSH attack, we find a BTI gadget in the default OpenSSH (version 7.2) SSH 
server binary available on Ubuntu 16.04 LTS, together with four \smother gadgets 
in \texttt{glibc} version 2.23, and leak bits of the host key. As shown by 
Heninger and Shacham~\cite{HeningerS09}, leaking a small 
fraction of bits enables recovery of the entire key.

The threat model for this attack assumes a local attacker that is able to 
initiate TCP connections to the ssh daemon. As before, we assume that ASLR is 
disabled (or can be bypassed through other means). Since the target BTI gadget 
runs pre-authentication, the attacker only needs to connect and does not need to 
authenticate to the server. In our PoC, the local attacker is running on the 
same host. However, the same attack can be run from a colocated VM, assuming the 
VMM schedules both attacker and victim VMs on the same physical core. We also 
assume that the attacker is able to spawn processes on the same physical core as 
the victim SSH process: the assumption is realistic, as shown for example by 
Zhang \textit{et al.}~\cite{ZhangJRR12}.

Our BTI gadget resides in the \Code{explicit_bzero} 
function~(\autoref{fig:real_bti_gadget_ssh}) which clears regions of memory. The 
function is extensively used to zero out sensitive data before memory is 
released as a countermeasure against data leakage if that memory region is 
reused for another purpose. To eliminate the possibility of dead-store 
optimization by the compiler, \Code{explicit_bzero} calls the standard 
\Code{bzero} function using a \emph{volatile} function pointer. We exploit the 
indirect jump generated for this function pointer call as the BTI gadget, 
knowing that the first argument to the function (stored in register \Code{rdi} 
according to the System V calling convention).

In particular, we exploit an invocation of the BTI gadget where the pointer 
refers to the server's private host key (e.g., RSA key). This invocation is 
present in the code path handling new connections, when the server loop forks 
new processes for each incoming connection and loads the private host key from 
disk with the \Code{key_load_private} function. The cryptographic values (e.g., 
the exponents and modulus of the RSA key) are kept in memory to later perform 
the ssh handshake but the buffer used to read out the file from disk is zeroed 
out and freed. This gadget is particularly convenient since the attacker gets
an arbitrary number of attempts at discovering different bits of 
the same private key. Also, the attacker can control when the victim process is 
spawned by initiating connections to the ssh daemon.

An abridged version of the \smother gadget is shown in 
\autoref{fig:real_victim_smother_ssh} (see \autoref{app:openssh_gadgets} for the 
full assembly listing). Our chosen \smother gadget differs slightly from that 
described in \autoref{sec:attack_principle} in that it compares the value of a 
memory location pointed to by a register, not the value of the register itself. 
The target and fall-through path differ in utilization of execution ports 0156.
This gadget is taken from \texttt{glibc} and demonstrates the 
availability of \smother-gadgets in commonly linked libraries. The attacker 
times a sequence of \Code{add} instructions with register operands (port 0156) 
shown in \autoref{fig:real_attacker-smother_ssh} to specifically target the same 
ports.

We ran our attack on a slightly-modified \Code{sshd} server. The ssh server is 
modified to setup relevant performance counters to be used for statistical and 
monitoring purposes. These counter values are ignored by the actual attack. 
The 
other modification is to synchronize the attacker with the BTI gadget (as in 
\autoref{sec:smotherspectrepoc}). 
For other targets (i.e., OpenSSL), we have investigated alternate
synchronization mechanisms that do not require victim modification and have good
results.
The server was compiled using the default options for Linux on x86\_64.

In the PoC of the attack, an orchestrator process randomly sets the bit to be 
leaked before launching the server and attacker on colocated logical cores. The 
attacker process is responsible for ``poisoning'' the BTB to cause 
mis-speculation on the victim process handling the incoming ssh connection. 
Prior to BTI, the attacker also performs a series of cache accesses that result 
in the eviction of the server's cache line containing the function pointer 
\Code{ssh_bzero}. This forces the victim's indirect call instruction to miss in 
the cache and speculate for a few cycles, increasing the BTI success rate. The 
attacker process is otherwise identical to the victim and follows the same code 
path, increasing the probability of the attacker having the same branching 
history as the victim at the call site, thereby increasing the success rate of 
BTI. The orchestrator launches an ssh client on a separate physical core to 
connect to the server and trigger the creation of the victim and attacker 
processes. Victim and attacker process execute and the attacker is able to 
collect a \smother-timing sample correlated to the value the LSB in byte 1 of 
the host private key.

The attack can be extended in two ways. First, we can pair our BTI gadget with 
other \smother gadgets in the victim, enabling us to leak other bits of the host 
private key. Second, we can find other occurrences of the 
\Code{explicit_bzero} BTI gadget (or other BTI gadgets) where different secrets 
are held in registers or in memory.

In the \Code{explicit_bzero} BTI gadget, we found that the value of the register 
\Code{r12} equals the value of \Code{rdi}, both pointing to the host key in 
memory at the point of attack. Therefore, we are able to use three other 
\smother-gadgets which dereference \Code{r12}. These gadgets allow us to leak 
extra bits from the host key, specifically the $4^{th}$ LSB in byte 13, the 
$4^{th}$ LSB in byte 14 and the $5^{th}$ LSB in byte 56. The corresponding 
assembly listings are shown in \aref{app:openssh_smother_r12}.

Additionally, we can also find other BTI gadgets, or invocations of 
\Code{explicit_bzero} with different secrets. Other secrets erased by this 
function include contents of the \Code{/etc/shadow} file and client passwords in 
cleartext received during login attempts.

\subsection{OpenSSL attack}
 \label{sec:opensslattack}

 \begin{figure*}[!t]
     \centering
     \begin{subfigure}[b]{.2\textwidth}
       \begin{minilisting}
 .rept 8;
 btrl r8d, r9d;
 btrl r10d, r11d;
 btsl r8d, r9d;
 btsl r10d, r11d;
 .endr;
       \end{minilisting}
       \caption{Attacker-timed code}
       \label{fig:real_attacker-smother}
     \end{subfigure}
     \begin{subfigure}[b]{.30\textwidth}
       \begin{minilisting}
 0xf5393:  testq  0x400,(rdx)
 0xf539a:  je     f5382 
 0xf539c:  mov    -0xb0(rbp),rdi
 ...                                    
 0xf5382:  add    0x1,rax
 ...     
       \end{minilisting}
       \caption{Victim {\smother} gadget (\texttt{glibc})}
       \label{fig:real_victim_smother}
     \end{subfigure}
     \begin{subfigure}[b]{.4\textwidth}
       \begin{minilisting}
 if(ctx->cipher->do_cipher(ctx, out, in, inl))
 {
     *outl = inl;
     return 1;
 } 
 ...
     \end{minilisting}
     \caption{Victim BTI gadget (OpenSSL)}
     \label{fig:real_bti_gadget}
   \end{subfigure}
   \caption{Gadgets from real-world libraries used 
           in our {\smotherspectre} exploit for leaking the $3^{rd}$ LSB of 
		byte 1 of \Code{rdx}'s pointee}
   \label{fig:real_smother_gadget_ssl}
 \end{figure*}

For OpenSSL, we target a BTI gadget in the
\Code{libcrypto} library (version 1.1.1b, dated 26-Feb-2019)  
which is widely used for performing 
cryptographic functions and a \smother gadget from \texttt{glibc} 
version 2.27.

Over the years, considerable effort was devoted to thwarting potential 
attackers and to protect OpenSSL from side-channel attacks, primarily by 
removing data-dependent memory-access or control flow. Our attack, however, 
targets BTI gadgets (indirect jumps or calls) that are found in code used to 
choose between encryption modes, allowing for multiple modes of operation (such
as ECB, CBC, GCM) with the same block cipher. 
OpenSSL uses a context variable that stores function pointers for
encryption/decryption. These pointers are set during the initialization phase
depending on the user-specified cipher mode.

As a result, cryptographic applications using \Code{libcrypto} execute an 
indirect call (the BTI gadget) during every block encryption or decryption. Such 
gadgets are the result of commonly used coding practices, and do not directly 
perform any data-dependent actions based on the secret value. As in the OpenSSH 
attack, the use of function pointers leads to indirect calls in the compiled 
binary. While security was the motivating factor for OpenSSH, OpenSSL uses 
function pointers to support polymorphic-like behavior, enabling our transient 
execution attack.

Our BTI gadget is contained in \texttt{EVP\_EncryptUpdate}, and is shown in 
\autoref{fig:real_bti_gadget}. The third argument~(\texttt{in}) contains a 
pointer to the plaintext to be encrypted (and is therefore a secret). In 
accordance with the System V calling convention, this pointer is stored in 
register \texttt{rdx} prior to the call. The secret in our chosen \smother 
gadget is the $3^{rd}$ LSB in byte 1 of the plaintext, referenced through 
\texttt{rdx}. An abridged version of the \smother gadget is shown in 
\autoref{fig:real_victim_smother} (see \autoref{app:openssl_gadgets} for the 
full assembly listing).

In our attack, we model a victim that encrypts text using OpenSSL's EnVeloP 
(EVP) API. After performing the necessary initializations, it performs a series 
of encryptions using calls to \texttt{EVP\_EncryptUpdate}. 
We have also instrumented the victim to setup relevant performance counters 
which are only used for statistical and monitoring purposes and are not used in 
the attack. The victim library does not contain any code to help the attacker 
synchronize with the execution of the BTI gadget. 

The attacker triggers the encryptions on the victim. It also runs code that is 
almost identical to the victim apart from the following differences. First, it 
loads the call pointer with the location of the \smother gadget on the victim to 
trigger BTI on the victim process. Second, it replaces the code at the target 
location by a delay sequence and the \smother timing. The delay sequence 
consists of a series of dependent instructions that allows the attacker to delay 
for a controlled number of cycles, synchronizing with the victim's \smother 
gadget, before measuring the timing sample. Otherwise, the attacker runs code 
that mimics the victim: it performs the same call to the encryption function 
where it follows the same sequence of checks and jumps. It also runs in a loop 
performing the same number of iterations, thus maximising the success of BTI.
In each iteration, the attacker gets one timing measurement. 
Between iterations, the attacker performs a series of memory accesses designed 
to evict the victim's cache line holding the pointer to the encryption function 
from the L1 cache to increase the BTI success rate. We observed that other usual 
work being performed on the core can have the same effect.

\subsection{Experimental results}

\begin{table}
  \centering
  {\sffamily\fontsize{8}{10}\selectfont
  \begin{tabulary}{\linewidth}{lrrr}
  \toprule
Pointer register & Byte Offset & Bit mask &  $\Delta$ \smother timing \\
  \midrule
    \Code{rdi}       & 0x01         & 0x01   & 0.32\% $\pm$ 0.21\% \\
    \Code{r12}       & 0x38         & 0x10   & 0.64\% $\pm$ 0.62\% \\
    \Code{r12}       & 0x0d         & 0x08   & 0.66\% $\pm$ 0.47\%  \\
    \Code{r12}       & 0x0c         & 0x08   & 0.42\% $\pm$ 0.33\%  \\
  \bottomrule
  \end{tabulary}
  }
  \caption{
\smotherspectre results leaking the \Code{sshd} private key. Four gadgets, each 
targeting a different key bit identified by its byte offset and bit mask, were 
used. We also show the mean timing difference percentage for the attacker's 
\smother timing, separated according to the value of the randomly-set target 
bit: all show a difference at 95\% confidence.
          }
  \label{tab:openssh_results}
\end{table}

We run the OpenSSH attack on a quad-core, hyper-threaded Skylake CPU (i7-6700K) 
with the server and attacker pinned on logical cores 0 and 4 respectively 
(running on physical core 0). For each connection attempt to the server, the 
orchestrator randomly sets or resets the target bit, logs its value and the 
attacker measures a \smotherspectre timing sample. We run the attack 10,000 times and 
separate the collected samples based on the value of the target bit on that 
particular run, yielding two sets of attacker timings corresponding to the 
target bit being zero or one. Finally, we run the Student's t-test to check 
whether the sets are statistically distinguishable. We used this methodology on 
four \smother-gadgets described in \autoref{sec:opensshattack}. 
\autoref{tab:openssh_results} shows the results: the distributions are 
differentiable with at-least 95\% confidence for each of the four gadgets. The 
attack does not require extremely high BTI success rate: in our samples, we 
observe BTI success rates ranging between 16\% and 25\%. The whole experiment 
takes about 75 seconds of real-time, of which a total of 20 seconds are spent by 
the orchestrator waiting for the server to be fully setup before launching the 
client.

\begin{figure}
    \includegraphics[width=\columnwidth]{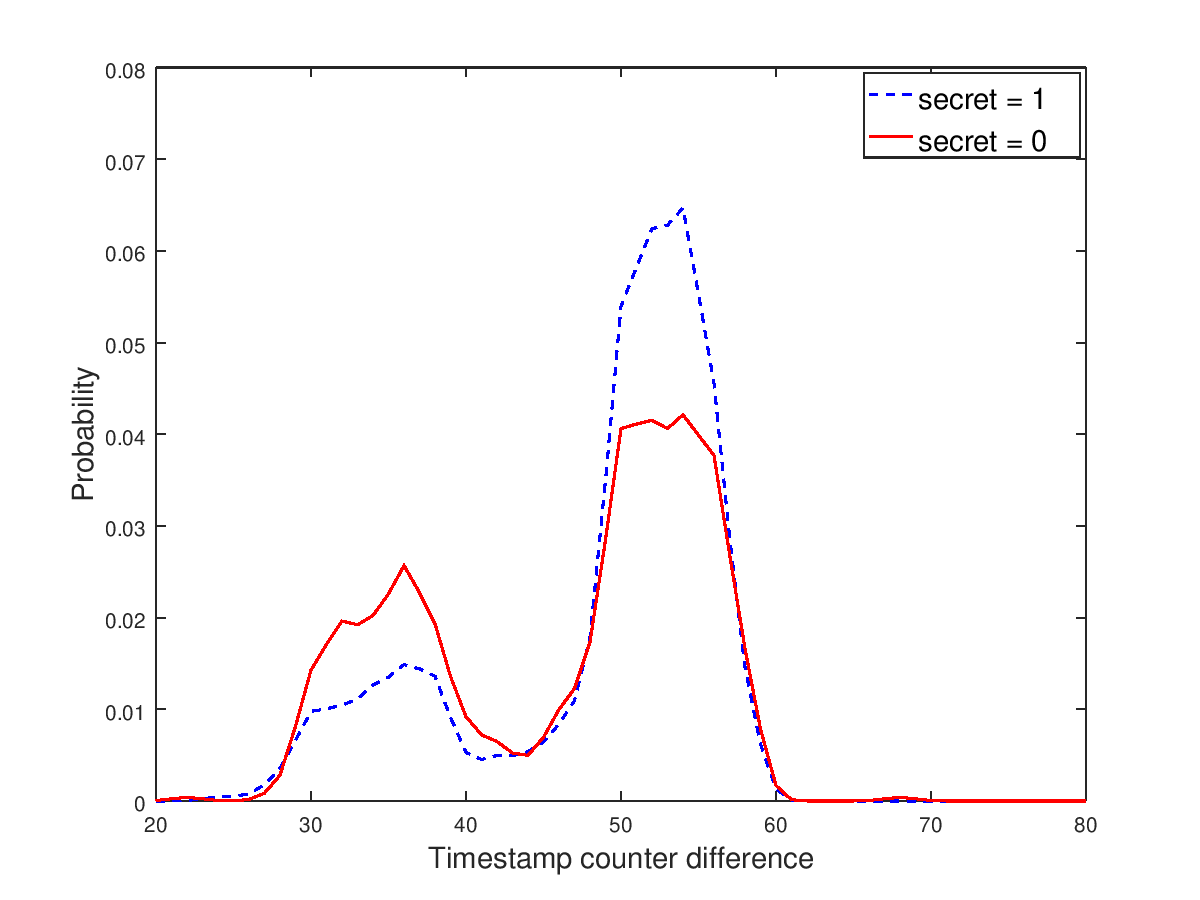}
    \caption{Probability density function (estimated using kernel density estimation) 
    for the attacker's timing running our
    \smotherspectre attack on OpenSSL, for when the victim's secret bit is one versus zero.}
    \label{fig:realattack_distributions}
\end{figure}

We run the OpenSSL attack on an i5-6200u CPU. A run of 100,000 encryptions is 
performed by the victim for each value of the secret bit. The large number of 
encryptions is necessary to estimate the probability density function for this 
\smother-gadget. A practical attack can confidently leak a bit with fewer 
encryptions. The attack takes about 950 ms of userspace time, leading to a 
measurement rate of more than 200,000 samples/second. The attacker succeeded in 
BTI with a success rate up to around $80\%$. We have found the time taken by the 
victim to reach the indirect call from the call triggering encryption entry to 
be highly predictable. The attacker is thus able to run the timing sequence 
concurrently with the victim's \smother-gadget without additional 
synchronization. \autoref{fig:realattack_distributions} shows the distribution 
of timestamp counter difference measured by the attacker for the \smother 
gadget. The distributions show a significant variation, with that corresponding 
to the zero-secret tending towards higher values. The Student's t-test is able 
to successfully distinguish between them with 95\% confidence. The test reports 
a timing difference of $10.69\% \pm 6.31\%$.

\subsection{Mitigating \smotherspectre}

Mitigations for \smotherspectre can be subdivided in two categories: mitigations
for \smother and mitigations for BTI. 

\paragraph{\smother mitigations}
%
The general idea of preventing \smother attacks is to ensure that two threads with
different privileges (in the general sense) do not compete for the same
execution port.

Currently available software \smother mitigations are limited. Apart from the
straightforward but performance-costly possibility of disabling SMT in its
entirety (up to 10-15\% overhead on Intel), the OS scheduler can employ a
side-channel aware strategy. For example, the OS scheduler can decide to only
colocate (on threads on the same core) processes from the same 
user~\cite{andersWranglingGhost}.

Finally, CPU-level mitigations could be deployed in the future, possibly
improving both security and performance over existing mitigations. For instance,
alternatives to SMT can be considered to achieve thread-level parallelism within
a core. These include coarse-grained and interleaved multithreading.

\paragraph{BTI mitigations}
%
Mitigations against branch target injection are also known as Spectre v2
mitigations. These include retpolines, which rewrite code to remove indirect
calls~\cite{turner2018retpoline}, as well as CPU-based controls.  Intel has 
exposed to
developers a set of security controls that limit an attacker's ability to
perform BTI. While they have been applied in selected cases, they have not been
widely adopted because of their overhead~\cite{corbetTamingStibp}, 
and because in many cases,
the required gadgets were simply not present~\cite{kocher18oakland}. 
Wide adoption of these
mitigations may limit the \smotherspectre attack.

\paragraph{Summary}
%
Fully mitigating the attack in either of these two categories is sufficient to
stop the attack presented in this paper. However, \smotherspectre does not
necessarily need to employ BTI: it can be generalized to use any other form of
speculative control flow hijack, e.g., Return Stack Buffer (RSB) 
overflow~\cite{mais18rsb} or speculative
return address overwrite~\cite{kiriansky18specoverflow}. In those cases, 
corresponding mitigations would apply.

\section{Related Work}
\label{sec:related}

\paragraph{Transient Execution Attacks}
%
Transient execution attacks exploit instructions that are executed, yet not
necessarily retired and thus cover both attacks based on speculative execution
as well as out-of-order execution~\cite{canella18arxiv}.

At the beginning of 2018, two security issues exploiting speculative execution
were revealed under the name ``Spectre''~\cite{kocher18oakland,
horn2018reading}. Spectre V1 (``Bounds Check Bypass'') exploits branch
prediction on a conditional branch to achieve an out-of-bounds access during
speculative execution: given a conditional branch that performs a bounds-check
on an array, the branch predictor is trained to the in-bounds case by
performing multiple executions of the corresponding code with a benign index.
When the code is then executed with an out-of-bounds index, a misprediction
occurs and the array access with the malicious index is performed. If the
result is used in further computation such as another array access, it can be
leaked through a side channel. Spectre V2 (``Branch Target Injection'')
exploits branch prediction on indirect control-flow transfers. To this end the
attacker first trains the branch predictor for a given address to transfer
control to an address of the attacker's choosing. The predictor will then use
the branch history created by the attacker for a spatially or temporally
co-located victim. Again, a cache side-channel can be used to leak data of the
attackers choosing in the following. The return stack buffer, which is used
for return statements in a similar fashion as the branch history is used for
indirect jumps has also been leveraged as a speculative execution
trigger~\cite{koruyeh18woot,mais18rsb}. The return address on the stack has
also been the target of other work, showing that through load-to-store
forwarding it can be speculatively overwritten, leading to a speculative
execution sibling of the classic stack buffer
overflow~\cite{kiriansky18specoverflow}.

Meltdown~\cite{lipp18sec} (``Rogue Data Cache Load''), which was also revealed
in early 2018 exploits out-of-order execution: a memory load instruction
immediately after a high latency instruction might fetch data into the cache
even if it is not permitted to access the actual memory location. The reason is
that on certain CPUs, the corresponding permission check is not on the critical
path for the data fetch and the exception is only triggered after the data
fetch. On such CPUs this allows reading arbitrary kernel memory from userspace.
Similarly, also privileged system register can be read (``Rogue System Register
Read''). The more recent Foreshadow~\cite{bulck18foreshadow} attacks a similar
phenomenon, ``L1 Terminal Fault'' in Intel nomenclature. If an instruction
accesses a virtual address that is not in the translation lookaside buffer (TLB)
and the corresponding page table entry's (PTE) present bit is not set, this is
referred to as a ``terminal fault''. During out-of-order execution, the
processor computes a physical address from the PTE, which is used for a lookup
in the L1 data cache. Until the instruction retires and a page fault is raised,
cached data is forwarded to dependent instructions, which can be used in an 
attack. This bypasses various access checks, including SGX protection, extended
page table address translation and system management mode (SMM) checks, thus
affecting virtualization and SGX enclaves (enclave data is not encrypted in
L1D). Also related to out-of-order execution is the speculative store
bypass~\cite{ms18ssb, amd18ssb}: for a code sequence of a dependent store and a load
instruction, the load instruction, if executed out-of-order before the store
might retrieve stale data from memory that can be used in a side channel. This
happens in cases where the CPU cannot detect the dependency in the code
sequence.

Transient execution attacks are not only a local security issue that requires a
victim device to execute attacker-controlled code. As
Netspectre~\cite{schwarz18arxiv} demonstrates they also work remotely. While
being less effective, they are still powerful enough to break, for example,
address space layout randomization.

\paragraph{Cache Side Channels}
Cache side channels leverage timing differences in accesses to different
tiers of the memory hierarchy.
Accesses to cached locations will be faster, whereas accesses to uncached
locations will be slower, as the data needs to be fetched from main memory.
This principle applies to both data and instructions:
Execution of code whose instructions are not cached will take longer than
execution of cached code.

To use an \textit{evict-and-time} cache side channel, one first primes the 
cache by executing a victim function and then measures how long the function
takes to execute -- this is the baseline run. One can now compare this
baseline against further executions of the function, with different cache sets
evicted. If the time the function takes to execute is slower than the
baseline, the victim function depends on the evicted cache set.

To use a \textit{prime-and-probe} cache side channel, one first primes the
cache with known attacker-controlled addresses. One then waits for the
victim code to run. Afterwards, one measures the access time to addresses
used for probing: it will be low for addresses touched by the victim code
and high for others. The difference to \textit{evict-and-time} is that the attacker
measures her own operation in contrast to the execution of victim code.
Both \textit{evict-and-time} and \textit{prime-and-probe} have been extensively
used to attack AES implementations~\cite{tromer05aes,tromer10aes}.

Another technique that became popular with attacks leveraging a shared
last-level cache (LLC) is \textit{flush-and-reload}. It requires an instruction
that allows an attacker to flush a certain cache line, such as \Code{clflush}
on x86\_64. In a corresponding attack, the attacker first flushes a cache line
and then waits for the victim code to execute. Afterwards the attacker times
the access to the address, which will be fast if the victim accessed (reloaded)
it and slow otherwise. Flush-and-reload is similar to prime-and-probe, but
much more fine-grained as individual cache lines can be targeted. It has been
used to leak information from the LLC, which is typically shared among multiple
CPU cores~\cite{yarom14}.
Related to flush-and-reload, \textit{flush-and-flush}~\cite{gruss16flush} is
based on the observation, that \Code{clflush} will take less time to execute
when it is run on a location that is not cached. The advantage over
flush-and-reload is that no actual access that would pull data into the cache
is performed, making the attack stealthier.

Finally, \textit{prime-and-abort} leverages Intel's transactional memory
mechanism to detect when a cache set has been evicted without the need to
probe the cache~\cite{disselkoen17abort}. In contrast to all previous cache
side channels, it does not need to time an operation. Transactional memory
operations require transactional data to be buffered in the cache which has
limited space. A transaction set up by the attacker will abort if the victim
accesses a critical address.

\paragraph{Other Side Channels}
%
Mitigations against cache-based side channels have led researchers to explore
other shared resources as well. TLBleed~\cite{gras18tlbleed} shows how the TLB
can be used as a side channel to leak a cryptographic key. Aforementioned
Netspectre-AVX~\cite{schwarz18arxiv} uses a side channel based on AVX
instructions. This side channel exploits the fact that the
execution unit processing those instructions employs aggressive power saving.
When such units have not been used for a long time, they execute much
slower. 

In particular, execution-unit-sharing-based side channels in the SMT settings
have been studied as early as in 2006: Wang and Lee~\cite{wang06smt} demonstrate
a multiply-based covert channel making use of contention on execution units.
Aciicmez and Seifert~\cite{aciicmez2007cheap} extend this work by analyzing its
applicability as a side channel. Anders Fogh~\cite{fogh16shotgun} proposes a
generalized result by analyzing contention results of the cross product of 12
curated instructions.  Finally, Portsmash~\cite{aldaya18portsmash}, concurrently
and independently demonstrates how port contention can be used to leak sensitive
cryptographic material from OpenSSL. Portsmash relies on a known vulnerable
implementation of OpenSSL, and therefore does not require any mitigation beyond
avoiding vulnerable code patterns. In contrast, \smotherspectre does not require a
secret-dependent control flow by combining port contention with BTI, and thereby
showing broader applicability of the port contention side channel.  Finally, in
contrast with all previous works, this work provides a characterization of this
side channel, including an analysis for low number of victim instructions.

\section{Conclusion}

We further our understanding of possible attacks in the space of speculative
execution. This is crucial to design suitable defenses and to apply them to the
right systems. In particular, we show that Branch Target Injection attacks
against applications that do not load attacker-provided code are feasible, by
crafting an exploit for the OpenSSH server and encryption using OpenSSL. 
To this end, we present a precise
characterisation of port contention, the non cache-based side channel we use for
the attack, and develop a tool to help us find suitable gadgets in existing
code. We will open-source our proof of concept implementation, gadget finder, as
well as the data of our measurements to enable others to study this interesting
side channel. As a consequence, it is now clear that in SMT environments
defenses solely relying on mitigating cache side channels, or solely relying on
reverting microarchitectural state after speculative execution, are
insufficient.

In the immediate future, implementing existing BTI mitigations is sufficient to
prevent \smotherspectre. Future work may mitigate such attacks with lower
performance overhead and better security guarantees, for instance through
side-channel resistant ways of designing thread-level parallelism in upcoming
CPUs.

\bibliographystyle{ACM-Reference-Format}
\bibliography{smotherspectre}


\begin{thebibliography}{37}


\ifx \showCODEN    \undefined \def \showCODEN     #1{\unskip}     \fi
\ifx \showDOI      \undefined \def \showDOI       #1{#1}\fi
\ifx \showISBNx    \undefined \def \showISBNx     #1{\unskip}     \fi
\ifx \showISBNxiii \undefined \def \showISBNxiii  #1{\unskip}     \fi
\ifx \showISSN     \undefined \def \showISSN      #1{\unskip}     \fi
\ifx \showLCCN     \undefined \def \showLCCN      #1{\unskip}     \fi
\ifx \shownote     \undefined \def \shownote      #1{#1}          \fi
\ifx \showarticletitle \undefined \def \showarticletitle #1{#1}   \fi
\ifx \showURL      \undefined \def \showURL       {\relax}        \fi
\providecommand\bibfield[2]{#2}
\providecommand\bibinfo[2]{#2}
\providecommand\natexlab[1]{#1}
\providecommand\showeprint[2][]{arXiv:#2}

\bibitem[\protect\citeauthoryear{Aciicmez and Seifert}{Aciicmez and
  Seifert}{2007}]%
        {aciicmez2007cheap}
\bibfield{author}{\bibinfo{person}{Onur Aciicmez} {and}
  \bibinfo{person}{Jean-Pierre Seifert}.} \bibinfo{year}{2007}\natexlab{}.
\newblock \showarticletitle{Cheap hardware parallelism implies cheap security}.
  In \bibinfo{booktitle}{\emph{Fault Diagnosis and Tolerance in Cryptography,
  2007. FDTC 2007. Workshop on}}. IEEE, \bibinfo{pages}{80--91}.
\newblock


\bibitem[\protect\citeauthoryear{Aldaya, Brumley, ul~Hassan, Garc\'ia, and
  Tuveri}{Aldaya et~al\mbox{.}}{2018}]%
        {aldaya18portsmash}
\bibfield{author}{\bibinfo{person}{Alejandro~Cabrera Aldaya},
  \bibinfo{person}{Billy~Bob Brumley}, \bibinfo{person}{Sohaib ul Hassan},
  \bibinfo{person}{Cesar~Pereida Garc\'ia}, {and} \bibinfo{person}{Nicola
  Tuveri}.} \bibinfo{year}{2018}\natexlab{}.
\newblock \bibinfo{title}{Port Contention for Fun and Profit}.
\newblock \bibinfo{howpublished}{Cryptology ePrint Archive, Report 2018/1060}.
\newblock
\newblock
\shownote{\url{https://eprint.iacr.org/2018/1060}.}


\bibitem[\protect\citeauthoryear{AMD}{AMD}{2018}]%
        {amd18ssb}
\bibfield{author}{\bibinfo{person}{AMD}.} \bibinfo{year}{2018}\natexlab{}.
\newblock \bibinfo{title}{Speculative Store Bypass Disable}.
\newblock
  \bibinfo{howpublished}{\url{https://developer.amd.com/wp-content/resources/124441_AMD64_SpeculativeStoreBypassDisable_Whitepaper_final.pdf}}.
\newblock


\bibitem[\protect\citeauthoryear{Bloom}{Bloom}{2018}]%
        {ccccloudsuite}
\bibfield{author}{\bibinfo{person}{Zack Bloom}.}
  \bibinfo{year}{2018}\natexlab{}.
\newblock \bibinfo{title}{Cloud Computing without Containers}.
\newblock
  \bibinfo{howpublished}{\url{https://blog.cloudflare.com/cloud-computing-without-containers/}}.
\newblock


\bibitem[\protect\citeauthoryear{Canella, Bulck, Schwarz, Lipp, von Berg,
  Ortner, Piessens, Evtyushkin, and Gruss}{Canella et~al\mbox{.}}{2018}]%
        {canella18arxiv}
\bibfield{author}{\bibinfo{person}{Claudio Canella}, \bibinfo{person}{Jo~Van
  Bulck}, \bibinfo{person}{Michael Schwarz}, \bibinfo{person}{Moritz Lipp},
  \bibinfo{person}{Benjamin von Berg}, \bibinfo{person}{Philipp Ortner},
  \bibinfo{person}{Frank Piessens}, \bibinfo{person}{Dmitry Evtyushkin}, {and}
  \bibinfo{person}{Daniel Gruss}.} \bibinfo{year}{2018}\natexlab{}.
\newblock \bibinfo{title}{A Systematic Evaluation of Transient Execution
  Attacks and Defenses}.
\newblock \bibinfo{howpublished}{\url{https://arxiv.org/abs/1811.05441}}.
\newblock


\bibitem[\protect\citeauthoryear{Coorporation}{Coorporation}{2016}]%
        {coorporation2009intel}
\bibfield{author}{\bibinfo{person}{Intel Coorporation}.}
  \bibinfo{year}{2016}\natexlab{}.
\newblock \bibinfo{title}{Intel 64 and IA-32 architectures optimization
  reference manual}.
\newblock
\newblock


\bibitem[\protect\citeauthoryear{Corbet}{Corbet}{[n.d.]}]%
        {corbetTamingStibp}
\bibfield{author}{\bibinfo{person}{Jonathan Corbet}.}
  \bibinfo{year}{[n.d.]}\natexlab{}.
\newblock \bibinfo{title}{Taming STIBP}.
\newblock \bibinfo{howpublished}{\url{ https://lwn.net/Articles/773118/ }}.
\newblock


\bibitem[\protect\citeauthoryear{Disselkoen, Kohlbrenner, Porter, and
  Tullsen}{Disselkoen et~al\mbox{.}}{2017}]%
        {disselkoen17abort}
\bibfield{author}{\bibinfo{person}{Craig Disselkoen}, \bibinfo{person}{David
  Kohlbrenner}, \bibinfo{person}{Leo Porter}, {and} \bibinfo{person}{Dean
  Tullsen}.} \bibinfo{year}{2017}\natexlab{}.
\newblock \showarticletitle{Prime+Abort: A Timer-Free High-Precision L3 Cache
  Attack using Intel {TSX}}. In \bibinfo{booktitle}{\emph{USENIX Security
  Symposium}}.
\newblock


\bibitem[\protect\citeauthoryear{Evtyushkin, Ponomarev, and
  Abu-Ghazaleh}{Evtyushkin et~al\mbox{.}}{2016}]%
        {evtyushkin2016jump}
\bibfield{author}{\bibinfo{person}{Dmitry Evtyushkin}, \bibinfo{person}{Dmitry
  Ponomarev}, {and} \bibinfo{person}{Nael Abu-Ghazaleh}.}
  \bibinfo{year}{2016}\natexlab{}.
\newblock \showarticletitle{Jump over ASLR: Attacking branch predictors to
  bypass ASLR}. In \bibinfo{booktitle}{\emph{The 49th Annual IEEE/ACM
  International Symposium on Microarchitecture}}. IEEE Press,
  \bibinfo{pages}{40}.
\newblock


\bibitem[\protect\citeauthoryear{Fog}{Fog}{[n.d.]}]%
        {fogInstTables}
\bibfield{author}{\bibinfo{person}{Agner Fog}.}
  \bibinfo{year}{[n.d.]}\natexlab{}.
\newblock \bibinfo{title}{{Instruction tables: Lists of instruction latencies,
  throughputs and micro-operation breakdowns for Intel, AMD and VIA CPUs}}.
\newblock
  \bibinfo{howpublished}{\url{https://www.agner.org/optimize/instruction_tables.pdf}}.
\newblock


\bibitem[\protect\citeauthoryear{Fogh}{Fogh}{[n.d.]}]%
        {fogh16shotgun}
\bibfield{author}{\bibinfo{person}{Anders Fogh}.}
  \bibinfo{year}{[n.d.]}\natexlab{}.
\newblock \bibinfo{title}{Covert Shotgun}.
\newblock \bibinfo{howpublished}{\url{
  https://cyber.wtf/2016/09/27/covert-shotgun/ }}.
\newblock


\bibitem[\protect\citeauthoryear{Fogh and Ertl}{Fogh and Ertl}{[n.d.]}]%
        {andersWranglingGhost}
\bibfield{author}{\bibinfo{person}{Anders Fogh} {and}
  \bibinfo{person}{Christopher Ertl}.} \bibinfo{year}{[n.d.]}\natexlab{}.
\newblock \bibinfo{title}{Wrangling with the Ghost: An inside story of
  mitigating speculative execution side channel vulnerabilities}.
\newblock
  \bibinfo{howpublished}{\url{https://i.blackhat.com/us-18/Thu-August-9/us-18-Fogh-Ertl-Wrangling-with-the-Ghost-An-Inside-Story-of-Mitigating-Speculative-Execution-Side-Channel-Vulnerabilities.pdf}}.
\newblock


\bibitem[\protect\citeauthoryear{Google}{Google}{[n.d.]}]%
        {gcloudfaq}
Google \bibinfo{year}{[n.d.]}\natexlab{}.
\newblock \bibinfo{title}{Google Compute Engine FAQ}.
\newblock
  \bibinfo{howpublished}{\url{https://cloud.google.com/compute/docs/faq}}.
\newblock
\newblock
\shownote{Accessed: 2019-02-13.}


\bibitem[\protect\citeauthoryear{Gras, Razavi, Bos, and Giuffrida}{Gras
  et~al\mbox{.}}{2018}]%
        {gras18tlbleed}
\bibfield{author}{\bibinfo{person}{Ben Gras}, \bibinfo{person}{Kaveh Razavi},
  \bibinfo{person}{Herbert Bos}, {and} \bibinfo{person}{Cristiano Giuffrida}.}
  \bibinfo{year}{2018}\natexlab{}.
\newblock \showarticletitle{Translation Leak-aside Buffer: Defeating Cache
  Side-channel Protections with {TLB} Attacks}. In
  \bibinfo{booktitle}{\emph{USENIX Security Symposium}}.
\newblock


\bibitem[\protect\citeauthoryear{Gruss, Lipp, Schwarz, Fellner, Maurice, and
  Mangard}{Gruss et~al\mbox{.}}{2017}]%
        {gruss2017kaslr}
\bibfield{author}{\bibinfo{person}{Daniel Gruss}, \bibinfo{person}{Moritz
  Lipp}, \bibinfo{person}{Michael Schwarz}, \bibinfo{person}{Richard Fellner},
  \bibinfo{person}{Cl{\'e}mentine Maurice}, {and} \bibinfo{person}{Stefan
  Mangard}.} \bibinfo{year}{2017}\natexlab{}.
\newblock \showarticletitle{Kaslr is dead: long live kaslr}. In
  \bibinfo{booktitle}{\emph{International Symposium on Engineering Secure
  Software and Systems}}. Springer, \bibinfo{pages}{161--176}.
\newblock


\bibitem[\protect\citeauthoryear{Gruss, Maurice, Wagner, and Mangard}{Gruss
  et~al\mbox{.}}{2016}]%
        {gruss16flush}
\bibfield{author}{\bibinfo{person}{Daniel Gruss},
  \bibinfo{person}{Cl{\'e}mentine Maurice}, \bibinfo{person}{Klaus Wagner},
  {and} \bibinfo{person}{Stefan Mangard}.} \bibinfo{year}{2016}\natexlab{}.
\newblock \showarticletitle{Flush+Flush: A Fast and Stealthy Cache Attack}. In
  \bibinfo{booktitle}{\emph{Detection of Intrusions and Malware, and
  Vulnerability Assessment}}.
\newblock


\bibitem[\protect\citeauthoryear{Heninger and Shacham}{Heninger and
  Shacham}{2009}]%
        {HeningerS09}
\bibfield{author}{\bibinfo{person}{Nadia Heninger} {and} \bibinfo{person}{Hovav
  Shacham}.} \bibinfo{year}{2009}\natexlab{}.
\newblock \showarticletitle{Reconstructing {RSA} Private Keys from Random Key
  Bits}. In \bibinfo{booktitle}{\emph{Advances in Cryptology - {CRYPTO} 2009,
  29th Annual International Cryptology Conference, Santa Barbara, CA, USA,
  August 16-20, 2009. Proceedings}} \emph{(\bibinfo{series}{Lecture Notes in
  Computer Science})}, \bibfield{editor}{\bibinfo{person}{Shai Halevi}} (Ed.),
  Vol.~\bibinfo{volume}{5677}. \bibinfo{publisher}{Springer},
  \bibinfo{pages}{1--17}.
\newblock
\showISBNx{978-3-642-03355-1}
\urldef\tempurl%
\url{https://doi.org/10.1007/978-3-642-03356-8\_1}
\showDOI{\tempurl}


\bibitem[\protect\citeauthoryear{Horn}{Horn}{2018}]%
        {horn2018reading}
\bibfield{author}{\bibinfo{person}{Jann Horn}.}
  \bibinfo{year}{2018}\natexlab{}.
\newblock \showarticletitle{Reading privileged memory with a side-channel}.
\newblock
  \bibinfo{howpublished}{\url{https://googleprojectzero.blogspot.com/2018/01/reading-privileged-memory-with-side.html}}.
\newblock \bibinfo{journal}{\emph{Project Zero}}  \bibinfo{volume}{3}
  (\bibinfo{year}{2018}).
\newblock


\bibitem[\protect\citeauthoryear{Hund, Willems, and Holz}{Hund
  et~al\mbox{.}}{2013}]%
        {hund2013practical}
\bibfield{author}{\bibinfo{person}{Ralf Hund}, \bibinfo{person}{Carsten
  Willems}, {and} \bibinfo{person}{Thorsten Holz}.}
  \bibinfo{year}{2013}\natexlab{}.
\newblock \showarticletitle{Practical timing side channel attacks against
  kernel space ASLR}. In \bibinfo{booktitle}{\emph{2013 IEEE Symposium on
  Security and Privacy}}. IEEE, \bibinfo{pages}{191--205}.
\newblock


\bibitem[\protect\citeauthoryear{Initiative}{Initiative}{2018}]%
        {ms18ssb}
\bibfield{author}{\bibinfo{person}{Secure~Windows Initiative}.}
  \bibinfo{year}{2018}\natexlab{}.
\newblock \bibinfo{title}{Speculative Store Bypass}.
\newblock
  \bibinfo{howpublished}{\url{https://blogs.technet.microsoft.com/srd/2018/05/21/analysis-and-mitigation-of-speculative-store-bypass-cve-2018-3639/}}.
\newblock


\bibitem[\protect\citeauthoryear{Khasawneh, Koruyeh, Song, Evtyushkin,
  Ponomarev, and Abu-Ghazaleh}{Khasawneh et~al\mbox{.}}{2018}]%
        {khasawneh2018safespec}
\bibfield{author}{\bibinfo{person}{Khaled~N Khasawneh},
  \bibinfo{person}{Esmaeil~Mohammadian Koruyeh}, \bibinfo{person}{Chengyu
  Song}, \bibinfo{person}{Dmitry Evtyushkin}, \bibinfo{person}{Dmitry
  Ponomarev}, {and} \bibinfo{person}{Nael Abu-Ghazaleh}.}
  \bibinfo{year}{2018}\natexlab{}.
\newblock \showarticletitle{SafeSpec: Banishing the Spectre of a Meltdown with
  Leakage-Free Speculation}.
\newblock \bibinfo{journal}{\emph{arXiv preprint arXiv:1806.05179}}
  (\bibinfo{year}{2018}).
\newblock


\bibitem[\protect\citeauthoryear{Kiriansky and Waldspurger}{Kiriansky and
  Waldspurger}{2018}]%
        {kiriansky18specoverflow}
\bibfield{author}{\bibinfo{person}{Vladimir Kiriansky} {and}
  \bibinfo{person}{Carl Waldspurger}.} \bibinfo{year}{2018}\natexlab{}.
\newblock \bibinfo{title}{{Speculative Buffer Overflows: Attacks and
  Defenses}}.
\newblock
  \bibinfo{howpublished}{\url{https://people.csail.mit.edu/vlk/spectre11.pdf}}.
\newblock


\bibitem[\protect\citeauthoryear{Kocher, Genkin, Gruss, Haas, Hamburg, Lipp,
  Mangard, Prescher, Schwarz, and Yarom}{Kocher et~al\mbox{.}}{2018}]%
        {kocher18oakland}
\bibfield{author}{\bibinfo{person}{Paul Kocher}, \bibinfo{person}{Daniel
  Genkin}, \bibinfo{person}{Daniel Gruss}, \bibinfo{person}{Werner Haas},
  \bibinfo{person}{Mike Hamburg}, \bibinfo{person}{Moritz Lipp},
  \bibinfo{person}{Stefan Mangard}, \bibinfo{person}{Thomas Prescher},
  \bibinfo{person}{Michael Schwarz}, {and} \bibinfo{person}{Yuval Yarom}.}
  \bibinfo{year}{2018}\natexlab{}.
\newblock \showarticletitle{Spectre Attacks: Exploiting Speculative Execution}.
  In \bibinfo{booktitle}{\emph{IEEE Symposium on Security and Privacy}}.
\newblock


\bibitem[\protect\citeauthoryear{Koruyeh, Khasawneh, Song, and
  Abu-Ghazaleh}{Koruyeh et~al\mbox{.}}{2018}]%
        {koruyeh18woot}
\bibfield{author}{\bibinfo{person}{Esmaeil~Mohammadian Koruyeh},
  \bibinfo{person}{Khaled~N. Khasawneh}, \bibinfo{person}{Chengyu Song}, {and}
  \bibinfo{person}{Nael Abu-Ghazaleh}.} \bibinfo{year}{2018}\natexlab{}.
\newblock \showarticletitle{Spectre Returns! Speculation Attacks using the
  Return Stack Buffer}. In \bibinfo{booktitle}{\emph{USENIX Workshop On
  Offensive Technologies}}.
\newblock


\bibitem[\protect\citeauthoryear{Laukemann, Hammer, Hofmann, Hager, and
  Wellein}{Laukemann et~al\mbox{.}}{2018}]%
        {laukemann18osaca}
\bibfield{author}{\bibinfo{person}{Jan Laukemann}, \bibinfo{person}{Julian
  Hammer}, \bibinfo{person}{Johannes Hofmann}, \bibinfo{person}{Georg Hager},
  {and} \bibinfo{person}{Gerhard Wellein}.} \bibinfo{year}{2018}\natexlab{}.
\newblock \bibinfo{title}{Automated Instruction Stream Throughput Prediction
  for Intel and AMD Microarchitectures}.
\newblock \bibinfo{howpublished}{\url{https://arxiv.org/abs/1809.00912}}.
\newblock


\bibitem[\protect\citeauthoryear{Lipp, Schwarz, Gruss, Prescher, Haas, Fogh,
  Horn, Mangard, Kocher, Genkin, Yarom, and Hamburg}{Lipp
  et~al\mbox{.}}{2018}]%
        {lipp18sec}
\bibfield{author}{\bibinfo{person}{Moritz Lipp}, \bibinfo{person}{Michael
  Schwarz}, \bibinfo{person}{Daniel Gruss}, \bibinfo{person}{Thomas Prescher},
  \bibinfo{person}{Werner Haas}, \bibinfo{person}{Anders Fogh},
  \bibinfo{person}{Jann Horn}, \bibinfo{person}{Stefan Mangard},
  \bibinfo{person}{Paul Kocher}, \bibinfo{person}{Daniel Genkin},
  \bibinfo{person}{Yuval Yarom}, {and} \bibinfo{person}{Mike Hamburg}.}
  \bibinfo{year}{2018}\natexlab{}.
\newblock \showarticletitle{Meltdown: Reading Kernel Memory from User Space}.
  In \bibinfo{booktitle}{\emph{USENIX Security Symposium}}.
\newblock


\bibitem[\protect\citeauthoryear{Maisuradze and Rossow}{Maisuradze and
  Rossow}{2018}]%
        {mais18rsb}
\bibfield{author}{\bibinfo{person}{Giorgi Maisuradze} {and}
  \bibinfo{person}{Christian Rossow}.} \bibinfo{year}{2018}\natexlab{}.
\newblock \showarticletitle{Ret2Spec: Speculative Execution Using Return Stack
  Buffers}. In \bibinfo{booktitle}{\emph{Conference on Computer and
  Communications Security}}.
\newblock


\bibitem[\protect\citeauthoryear{Osvik, Shamir, and Tromer}{Osvik
  et~al\mbox{.}}{2006}]%
        {tromer05aes}
\bibfield{author}{\bibinfo{person}{Dag~Arne Osvik}, \bibinfo{person}{Adi
  Shamir}, {and} \bibinfo{person}{Eran Tromer}.}
  \bibinfo{year}{2006}\natexlab{}.
\newblock \showarticletitle{Cache Attacks and Countermeasures: The Case of
  AES}. In \bibinfo{booktitle}{\emph{Topics in Cryptology}}.
\newblock


\bibitem[\protect\citeauthoryear{Schwarz, Schwarzl, Lipp, and Gruss}{Schwarz
  et~al\mbox{.}}{2018}]%
        {schwarz18arxiv}
\bibfield{author}{\bibinfo{person}{Michael Schwarz}, \bibinfo{person}{Martin
  Schwarzl}, \bibinfo{person}{Moritz Lipp}, {and} \bibinfo{person}{Daniel
  Gruss}.} \bibinfo{year}{2018}\natexlab{}.
\newblock \bibinfo{title}{NetSpectre: Read Arbitrary Memory over Network}.
\newblock \bibinfo{howpublished}{\url{https://arxiv.org/abs/1807.10535}}.
\newblock


\bibitem[\protect\citeauthoryear{Sotirov}{Sotirov}{2009}]%
        {sotirov2009bypassing}
\bibfield{author}{\bibinfo{person}{Alexander Sotirov}.}
  \bibinfo{year}{2009}\natexlab{}.
\newblock \showarticletitle{Bypassing memory protections: The future of
  exploitation}. In \bibinfo{booktitle}{\emph{USENIX Security}}.
\newblock


\bibitem[\protect\citeauthoryear{Torvalds}{Torvalds}{2018}]%
        {linusemails}
\bibfield{author}{\bibinfo{person}{Linus Torvalds}.}
  \bibinfo{year}{2018}\natexlab{}.
\newblock \bibinfo{title}{Linus on Spectre/Meltdown mitigations}.
\newblock \bibinfo{howpublished}{\url{https://lkml.org/lkml/2018/1/21/192}}.
\newblock


\bibitem[\protect\citeauthoryear{Tromer, Osvik, and Shamir}{Tromer
  et~al\mbox{.}}{2010}]%
        {tromer10aes}
\bibfield{author}{\bibinfo{person}{Eran Tromer}, \bibinfo{person}{Dag~Arne
  Osvik}, {and} \bibinfo{person}{Adi Shamir}.} \bibinfo{year}{2010}\natexlab{}.
\newblock \showarticletitle{Efficient Cache Attacks on AES, and
  Countermeasures}.
\newblock \bibinfo{journal}{\emph{Journal of Cryptology}}
  (\bibinfo{year}{2010}).
\newblock


\bibitem[\protect\citeauthoryear{Turner}{Turner}{2018}]%
        {turner2018retpoline}
\bibfield{author}{\bibinfo{person}{Paul Turner}.}
  \bibinfo{year}{2018}\natexlab{}.
\newblock \bibinfo{title}{Retpoline: a software construct for preventing
  branch-target-injection}.
\newblock
  \bibinfo{howpublished}{\url{https://support.google.com/faqs/answer/7625886}}.
\newblock


\bibitem[\protect\citeauthoryear{Van~Bulck, Minkin, Weisse, Genkin, Kasikci,
  Piessens, Silberstein, Wenisch, Yarom, and Strackx}{Van~Bulck
  et~al\mbox{.}}{2018}]%
        {bulck18foreshadow}
\bibfield{author}{\bibinfo{person}{Jo Van~Bulck}, \bibinfo{person}{Marina
  Minkin}, \bibinfo{person}{Ofir Weisse}, \bibinfo{person}{Daniel Genkin},
  \bibinfo{person}{Baris Kasikci}, \bibinfo{person}{Frank Piessens},
  \bibinfo{person}{Mark Silberstein}, \bibinfo{person}{Thomas~F. Wenisch},
  \bibinfo{person}{Yuval Yarom}, {and} \bibinfo{person}{Raoul Strackx}.}
  \bibinfo{year}{2018}\natexlab{}.
\newblock \showarticletitle{Foreshadow: Extracting the Keys to the {Intel SGX}
  Kingdom with Transient Out-of-Order Execution}. In
  \bibinfo{booktitle}{\emph{USENIX Security Symposium}}.
\newblock


\bibitem[\protect\citeauthoryear{Wang and Lee}{Wang and Lee}{2006}]%
        {wang06smt}
\bibfield{author}{\bibinfo{person}{Zhenghong Wang} {and}
  \bibinfo{person}{Ruby~B. Lee}.} \bibinfo{year}{2006}\natexlab{}.
\newblock \showarticletitle{Covert and Side Channels Due to Processor
  Architecture}. In \bibinfo{booktitle}{\emph{Proceedings of the 22Nd Annual
  Computer Security Applications Conference}} \emph{(\bibinfo{series}{ACSAC
  '06})}. \bibinfo{publisher}{IEEE Computer Society},
  \bibinfo{address}{Washington, DC, USA}, \bibinfo{pages}{473--482}.
\newblock
\showISBNx{0-7695-2716-7}
\urldef\tempurl%
\url{https://doi.org/10.1109/ACSAC.2006.20}
\showDOI{\tempurl}


\bibitem[\protect\citeauthoryear{Yarom and Falkner}{Yarom and Falkner}{2014}]%
        {yarom14}
\bibfield{author}{\bibinfo{person}{Yuval Yarom} {and} \bibinfo{person}{Katrina
  Falkner}.} \bibinfo{year}{2014}\natexlab{}.
\newblock \showarticletitle{{FLUSH+RELOAD: A High Resolution, Low Noise, L3
  Cache Side-channel Attack}}. In \bibinfo{booktitle}{\emph{USENIX Security
  Symposium}}.
\newblock


\bibitem[\protect\citeauthoryear{Zhang, Juels, Reiter, and Ristenpart}{Zhang
  et~al\mbox{.}}{2012}]%
        {ZhangJRR12}
\bibfield{author}{\bibinfo{person}{Yinqian Zhang}, \bibinfo{person}{Ari Juels},
  \bibinfo{person}{Michael~K. Reiter}, {and} \bibinfo{person}{Thomas
  Ristenpart}.} \bibinfo{year}{2012}\natexlab{}.
\newblock \showarticletitle{Cross-VM side channels and their use to extract
  private keys}. In \bibinfo{booktitle}{\emph{the {ACM} Conference on Computer
  and Communications Security, CCS'12, Raleigh, NC, USA, October 16-18, 2012}},
  \bibfield{editor}{\bibinfo{person}{Ting Yu}, \bibinfo{person}{George
  Danezis}, {and} \bibinfo{person}{Virgil~D. Gligor}} (Eds.).
  \bibinfo{publisher}{{ACM}}, \bibinfo{pages}{305--316}.
\newblock
\showISBNx{978-1-4503-1651-4}
\urldef\tempurl%
\url{https://doi.org/10.1145/2382196.2382230}
\showDOI{\tempurl}


\end{thebibliography}

\appendix
\section{Gadgets leaking 21 bits of \texttt{rax}}
\label{app:rax_gadgets}

The following table lists parts of \smother-gadgets which can be used to leak 
21 bits of information from \Code{rax}. We also show which library the gadget
was found in.

  \begin{center}
  {\sffamily\fontsize{8}{10}\selectfont}
  \begin{tabulary}{\linewidth}{Lll}
  \toprule
  Address & Comparison instruction & Library \\
  \midrule
0xd3900   & test  0x1, al           & glibc \\
0x1101cb  & test  0x2, al           & glibc \\
0x12f779  & and   0x4, al           & glibc \\
0x29709   & and   0x8, al           & glibc   \\
0x126500  & and   0x10, al          & glibc   \\
0x7e83    & and   0x20, al          & ld   \\
0xc378e   & and   0x40, al          & glibc   \\
0xd7e50   & and   0x80, eax         & glibc   \\
0x12cad9  & test  0x2, ah           & stdc++   \\
0xf1794   & test  0x307, ax         & libcrypto   \\
0x5f661   & and   0x2100, eax       & glibc   \\
0x11c7f6  & and   0x2abd, eax       & glibc   \\
0x10ca11  & and   0x8000, eax       & glibc \\
0x17bcd4  & test  0x100000, eax     & libcrypto  \\
0x268de   & test  0x200000, eax     & ssl   \\
0xbe656   & and   0x3084a5, eax     & glibc   \\
0x26f20   & test  0x800000, eax     & ssl   \\
0xb3ba0   & test  0x1000000, eax    & glibc   \\
0xb7db    & test  0x40000000, eax   & pthread  \\
0x50e7b   & test  0x80000000, eax   & ssl   \\
0xa6133   & test  0x83000002, eax   & libcrypto   \\
  \bottomrule
  \end{tabulary}
\end{center}

\section{OpenSSL attack gadgets}
\label{app:openssl_gadgets}

\subsection{\smother gadget}
\label{app:openssl_smother}

The following gadget leaks the $3^{rd}$ LSB from the byte
at offset 1 from the pointer in \Code{rdx}.
\begin{verbatim}
f5393:  testq  0x400, (rdx)
f539a:  je     f5382 
f539c:  mov    -0xb0(rbp), rdi
f53a3:  mov    -0xf0(rbp), edx
f53a9:  mov    (rdi, rax, 8), rax
f53ad:  test   edx, edx
f53af:  mov    rax, 0x50(rbx)
...
f5382:  add    0x1, rax
f5386:  add    0x20, rdx
f538a:  cmp    rax, -0x100(rbp)
...
\end{verbatim}

\section{OpenSSH attack gadgets}
\label{app:openssh_gadgets}

\subsection{\smother gadget with \Code{rdi} pointer}
\label{app:openssh_smother_rdi}

The following gadget leaks the LSB from the byte
at offset 1 from the pointer in \Code{rdi}.
\begin{verbatim}   
...
6f8dc:  testl  0x100, (rdi)
6f8e2:  je     6f8ef 
6f8e4:  mov    0x10(rbx), rax
6f8e8:  sub    0x8(rbx), rax
6f8ec:  sub    rax, rsi
6f8ef:  mov    rbx, rdi
6f8f2:  mov    ecx, 0xc(rsp)
6f8f6:  mov    edx, 0x8(rsp)
6f8fa:  mov    rsi, (rsp)
...
\end{verbatim}

\subsection{\smother gadgets with \Code{r12} pointer}
\label{app:openssh_smother_r12}

The following gadget leaks the $5^{th}$ LSB from the byte
at offset 56 from the pointer in \Code{r12}.
\begin{verbatim}
e8577:  testb  0x10, 0x38(r12)
e857d:  je     e8608 
e8583:  sub    0x8, rsp
e8587:  push   rbx
e8588:  pushq  0x0
e858a:  pushq  0x0
e858c:  mov    edx, r8d
e858f:  mov    edx, r9d
e8592:  mov    r10d, ecx
e8595:  sub    r10d, r8d
e8598:  mov    r13, rsi
e859b:  mov    r12, rdi
...   
e8608:  sub    0x8,rsp
e860c:  push   rbx
e860d:  push   r14
e860f:  push   r15
...
\end{verbatim}

The following gadget leaks the $4^{th}$ LSB from the byte
at offset 12 from the pointer in \Code{r12}.
\begin{verbatim}   
5220e:  testb  0x8, 0xd(r12)
52214:  je     52221 
52216:  xor    edx, edx
52218:  xor    esi, esi
5221a:  xor    edi, edi
...
52221:  mov    r13, rcx
52224:  add    0x3, r13
52228:  sar    0x2, rcx
5222c:  cmp    0x6, r13
...
\end{verbatim}

The following gadget leaks the $4^{th}$ LSB from the byte
at offset 13 from the pointer in \Code{r12}.
\begin{verbatim}
529a2:  testb  0x8, 0xc(r12)
529a8:  je     523da 
529ae:  mov    -0x100(rbp), rcx
529b5:  lea    0xc(rcx), rdx
529b9:  cmp    rax, rdx
...
523da:  mov    -0xf8(rbp), r13d
523e1:  mov    (rax), edx
523e3:  add    -0xe8(rbp), r13d
...
\end{verbatim}

\section{Responsible disclosure}
The attacks presented in this paper were disclosed to Intel, OpenSSL and AMD in
late 2018.

\newcommand\blfootnote[1]{%
    \begingroup
      \renewcommand\thefootnote{}\footnote{#1}%
        \addtocounter{footnote}{-1}%
          \endgroup
        }

        \blfootnote{
\copyright Copyright International Business Machines Corporation and EPFL 2019 \\
All Rights Reserved\\
Printed in the United States of America (09/19/2019)\\
The following are trademarks of International Business Machines Corporation in
the United States, or other countries, or both.\\
IBM\\
IBM Research\\
IBM Z\\
POWER\\

Other company, product, and service names may be trademarks or service marks of
others.  All information contained in this document is subject to change without
notice.  The products described in this document are NOT intended for use in
implantation, life support, space, nuclear, or military applications where
malfunction may result in injury or death to persons. The information contained
in this document does not affect or change IBM product specifications or
warranties. Nothing in this document shall operate as an express or implied
license or indemnity under the intellectual property rights of IBM or third
parties. All information contained in this document was obtained in specific
environments, and is presented as an illustration. The results obtained in other
operating environments may vary.  THE INFORMATION CONTAINED IN THIS DOCUMENT IS
PROVIDED ON AN "AS IS" BASIS. In no event will IBM be liable for damages arising
directly or indirectly from any use of the information contained in this
document.\\
IBM Corporation\\
New Orchard Road\\
Armonk, NY 10504
}

\end{document}